\documentclass[longauth]{aaEC}
\usepackage{lastpage}
\usepackage{graphicx}
\usepackage{natbib}
\usepackage{scalerel}
\usepackage[table]{xcolor}
\usepackage{rotating}
\usepackage{amsmath}

\newcommand{\Mstarsun}{\logten(M_{\ast}/M_\odot)}
\newcommand{\Mstar}{M_{\ast}}
\defcitealias{Q1-SP003}{Euclid Collaboration: Roster et al. 2025}

\usepackage{txfonts}
\usepackage[pdfencoding=auto,psdextra]{hyperref}
\hypersetup{
    colorlinks=true,
    linkcolor=blue,
    filecolor=magenta,      
    urlcolor=blue,
    citecolor=blue
}
\usepackage[nameinlink,capitalise]{cleveref}
\crefname{section}{Sect.}{Sects.}
\Crefname{section}{Section}{Sections}
\crefname{figure}{Fig.}{Figs.}
\Crefname{figure}{Figure}{Figures}
\crefname{equation}{Eq.}{Eqs.}
\Crefname{equation}{Equation}{Equations}
\crefname{table}{Table}{Tables}
\crefname{appendix}{Appendix}{Appendices}
\creflabelformat{equation}{#2\textup{#1}#3}

\makeatletter
\renewcommand*\aa@pageof{, page \thepage{} of \pageref*{LastPage}}
\makeatother

\usepackage[utf8]{inputenc}

\usepackage[switch, modulo]{lineno}

\usepackage{euclid}

\defcitealias{Q1-SP027}{EC:MZ25}

\begin{document}
%
%

\title{\Euclid Quick Data Release (Q1): Dual active galactic nuclei in low-mass galaxies\thanks{This paper is published on behalf of the Euclid Consortium.}}    


\newcommand{\orcid}[1]{} 
\author{M.~Mezcua\orcid{0000-0003-4440-259X}\thanks{\email{mezcua@ice.csic.es}}\inst{\ref{aff1},\ref{aff2}}
\and B.~Laloux\orcid{0000-0001-9996-9732}\inst{\ref{aff3},\ref{aff4}}
\and M.~Scialpi\orcid{0009-0006-5100-4986}\inst{\ref{aff5},\ref{aff6},\ref{aff7}}
\and M.~Siudek\orcid{0000-0002-2949-2155}\inst{\ref{aff1},\ref{aff8},\ref{aff9}}
\and A.~Er\'ostegui\orcid{0000-0002-6729-2373}\inst{\ref{aff10}}
\and F.~Ricci\orcid{0000-0001-5742-5980}\inst{\ref{aff11},\ref{aff12}}
\and T.~Matamoro~Zatarain\orcid{0009-0007-2976-293X}\inst{\ref{aff13}}
\and S.~Visser\orcid{0009-0004-4241-4911}\inst{\ref{aff14}}
\and H.~J.~A.~Rottgering\orcid{0000-0001-8887-2257}\inst{\ref{aff14}}
\and C.~M.~Gutierrez\orcid{0000-0001-7854-783X}\inst{\ref{aff8},\ref{aff9}}
\and A.~Feltre\orcid{0000-0001-6865-2871}\inst{\ref{aff7}}
\and L.~Bisigello\orcid{0000-0003-0492-4924}\inst{\ref{aff15}}
\and C.~Saulder\orcid{0000-0002-0408-5633}\inst{\ref{aff4},\ref{aff16}}
\and L.~Ulivi\orcid{0009-0001-3291-5382}\inst{\ref{aff6},\ref{aff5},\ref{aff7}}
\and J.~H.~Knapen\orcid{0000-0003-1643-0024}\inst{\ref{aff8},\ref{aff9}}
\and H.~Dom\'inguez~S\'anchez\orcid{0000-0002-9013-1316}\inst{\ref{aff17}}
\and G.~Zamorani\orcid{0000-0002-2318-301X}\inst{\ref{aff18}}
\and K.~Rubinur\orcid{0000-0001-5574-5104}\inst{\ref{aff19}}
\and J.~Calhau\orcid{0000-0003-1803-6899}\inst{\ref{aff3}}
\and L.~Spinoglio\orcid{0000-0001-8840-1551}\inst{\ref{aff20}}
\and F.~Shankar\orcid{0000-0001-8973-5051}\inst{\ref{aff21}}
\and D.~Stern\orcid{0000-0003-2686-9241}\inst{\ref{aff22}}
\and R.~Pucha\orcid{0000-0002-4940-3009}\inst{\ref{aff23}}
\and A.~Viitanen\orcid{0000-0001-9383-786X}\inst{\ref{aff24},\ref{aff25},\ref{aff12}}
\and B.~Altieri\orcid{0000-0003-3936-0284}\inst{\ref{aff26}}
\and S.~Andreon\orcid{0000-0002-2041-8784}\inst{\ref{aff27}}
\and N.~Auricchio\orcid{0000-0003-4444-8651}\inst{\ref{aff18}}
\and M.~Baldi\orcid{0000-0003-4145-1943}\inst{\ref{aff28},\ref{aff18},\ref{aff29}}
\and S.~Bardelli\orcid{0000-0002-8900-0298}\inst{\ref{aff18}}
\and P.~Battaglia\orcid{0000-0002-7337-5909}\inst{\ref{aff18}}
\and A.~Biviano\orcid{0000-0002-0857-0732}\inst{\ref{aff30},\ref{aff31}}
\and M.~Brescia\orcid{0000-0001-9506-5680}\inst{\ref{aff32},\ref{aff3}}
\and S.~Camera\orcid{0000-0003-3399-3574}\inst{\ref{aff33},\ref{aff34},\ref{aff35}}
\and G.~Ca\~nas-Herrera\orcid{0000-0003-2796-2149}\inst{\ref{aff36},\ref{aff14}}
\and V.~Capobianco\orcid{0000-0002-3309-7692}\inst{\ref{aff35}}
\and C.~Carbone\orcid{0000-0003-0125-3563}\inst{\ref{aff37}}
\and J.~Carretero\orcid{0000-0002-3130-0204}\inst{\ref{aff38},\ref{aff39}}
\and M.~Castellano\orcid{0000-0001-9875-8263}\inst{\ref{aff12}}
\and G.~Castignani\orcid{0000-0001-6831-0687}\inst{\ref{aff18}}
\and S.~Cavuoti\orcid{0000-0002-3787-4196}\inst{\ref{aff3},\ref{aff40}}
\and K.~C.~Chambers\orcid{0000-0001-6965-7789}\inst{\ref{aff41}}
\and A.~Cimatti\inst{\ref{aff42}}
\and C.~Colodro-Conde\inst{\ref{aff8}}
\and G.~Congedo\orcid{0000-0003-2508-0046}\inst{\ref{aff36}}
\and C.~J.~Conselice\orcid{0000-0003-1949-7638}\inst{\ref{aff43}}
\and L.~Conversi\orcid{0000-0002-6710-8476}\inst{\ref{aff44},\ref{aff26}}
\and Y.~Copin\orcid{0000-0002-5317-7518}\inst{\ref{aff45}}
\and F.~Courbin\orcid{0000-0003-0758-6510}\inst{\ref{aff46},\ref{aff47},\ref{aff10}}
\and H.~M.~Courtois\orcid{0000-0003-0509-1776}\inst{\ref{aff48}}
\and M.~Cropper\orcid{0000-0003-4571-9468}\inst{\ref{aff49}}
\and H.~Degaudenzi\orcid{0000-0002-5887-6799}\inst{\ref{aff25}}
\and G.~De~Lucia\orcid{0000-0002-6220-9104}\inst{\ref{aff30}}
\and C.~Dolding\orcid{0009-0003-7199-6108}\inst{\ref{aff49}}
\and H.~Dole\orcid{0000-0002-9767-3839}\inst{\ref{aff50}}
\and F.~Dubath\orcid{0000-0002-6533-2810}\inst{\ref{aff25}}
\and X.~Dupac\inst{\ref{aff26}}
\and M.~Farina\orcid{0000-0002-3089-7846}\inst{\ref{aff20}}
\and R.~Farinelli\inst{\ref{aff18}}
\and F.~Faustini\orcid{0000-0001-6274-5145}\inst{\ref{aff12},\ref{aff51}}
\and S.~Ferriol\inst{\ref{aff45}}
\and M.~Frailis\orcid{0000-0002-7400-2135}\inst{\ref{aff30}}
\and E.~Franceschi\orcid{0000-0002-0585-6591}\inst{\ref{aff18}}
\and M.~Fumana\orcid{0000-0001-6787-5950}\inst{\ref{aff37}}
\and S.~Galeotta\orcid{0000-0002-3748-5115}\inst{\ref{aff30}}
\and K.~George\orcid{0000-0002-1734-8455}\inst{\ref{aff52}}
\and B.~Gillis\orcid{0000-0002-4478-1270}\inst{\ref{aff36}}
\and C.~Giocoli\orcid{0000-0002-9590-7961}\inst{\ref{aff18},\ref{aff29}}
\and J.~Gracia-Carpio\inst{\ref{aff4}}
\and A.~Grazian\orcid{0000-0002-5688-0663}\inst{\ref{aff15}}
\and F.~Grupp\inst{\ref{aff4},\ref{aff16}}
\and S.~V.~H.~Haugan\orcid{0000-0001-9648-7260}\inst{\ref{aff19}}
\and J.~Hoar\inst{\ref{aff26}}
\and H.~Hoekstra\orcid{0000-0002-0641-3231}\inst{\ref{aff14}}
\and W.~Holmes\inst{\ref{aff22}}
\and I.~M.~Hook\orcid{0000-0002-2960-978X}\inst{\ref{aff53}}
\and F.~Hormuth\inst{\ref{aff54}}
\and A.~Hornstrup\orcid{0000-0002-3363-0936}\inst{\ref{aff55},\ref{aff56}}
\and K.~Jahnke\orcid{0000-0003-3804-2137}\inst{\ref{aff57}}
\and M.~Jhabvala\inst{\ref{aff58}}
\and S.~Kermiche\orcid{0000-0002-0302-5735}\inst{\ref{aff59}}
\and B.~Kubik\orcid{0009-0006-5823-4880}\inst{\ref{aff45}}
\and M.~K\"ummel\orcid{0000-0003-2791-2117}\inst{\ref{aff16}}
\and M.~Kunz\orcid{0000-0002-3052-7394}\inst{\ref{aff60}}
\and H.~Kurki-Suonio\orcid{0000-0002-4618-3063}\inst{\ref{aff61},\ref{aff62}}
\and A.~M.~C.~Le~Brun\orcid{0000-0002-0936-4594}\inst{\ref{aff63}}
\and S.~Ligori\orcid{0000-0003-4172-4606}\inst{\ref{aff35}}
\and P.~B.~Lilje\orcid{0000-0003-4324-7794}\inst{\ref{aff19}}
\and V.~Lindholm\orcid{0000-0003-2317-5471}\inst{\ref{aff61},\ref{aff62}}
\and I.~Lloro\orcid{0000-0001-5966-1434}\inst{\ref{aff64}}
\and G.~Mainetti\orcid{0000-0003-2384-2377}\inst{\ref{aff65}}
\and D.~Maino\inst{\ref{aff66},\ref{aff37},\ref{aff67}}
\and E.~Maiorano\orcid{0000-0003-2593-4355}\inst{\ref{aff18}}
\and O.~Mansutti\orcid{0000-0001-5758-4658}\inst{\ref{aff30}}
\and O.~Marggraf\orcid{0000-0001-7242-3852}\inst{\ref{aff68}}
\and M.~Martinelli\orcid{0000-0002-6943-7732}\inst{\ref{aff12},\ref{aff69}}
\and N.~Martinet\orcid{0000-0003-2786-7790}\inst{\ref{aff70}}
\and F.~Marulli\orcid{0000-0002-8850-0303}\inst{\ref{aff71},\ref{aff18},\ref{aff29}}
\and R.~J.~Massey\orcid{0000-0002-6085-3780}\inst{\ref{aff72}}
\and E.~Medinaceli\orcid{0000-0002-4040-7783}\inst{\ref{aff18}}
\and S.~Mei\orcid{0000-0002-2849-559X}\inst{\ref{aff73},\ref{aff74}}
\and Y.~Mellier\thanks{Deceased}\inst{\ref{aff75},\ref{aff76}}
\and M.~Meneghetti\orcid{0000-0003-1225-7084}\inst{\ref{aff18},\ref{aff29}}
\and E.~Merlin\orcid{0000-0001-6870-8900}\inst{\ref{aff12}}
\and G.~Meylan\inst{\ref{aff77}}
\and A.~Mora\orcid{0000-0002-1922-8529}\inst{\ref{aff78}}
\and M.~Moresco\orcid{0000-0002-7616-7136}\inst{\ref{aff71},\ref{aff18}}
\and L.~Moscardini\orcid{0000-0002-3473-6716}\inst{\ref{aff71},\ref{aff18},\ref{aff29}}
\and C.~Neissner\orcid{0000-0001-8524-4968}\inst{\ref{aff79},\ref{aff39}}
\and R.~C.~Nichol\orcid{0000-0003-0939-6518}\inst{\ref{aff80}}
\and S.-M.~Niemi\orcid{0009-0005-0247-0086}\inst{\ref{aff81}}
\and C.~Padilla\orcid{0000-0001-7951-0166}\inst{\ref{aff79}}
\and S.~Paltani\orcid{0000-0002-8108-9179}\inst{\ref{aff25}}
\and F.~Pasian\orcid{0000-0002-4869-3227}\inst{\ref{aff30}}
\and K.~Pedersen\inst{\ref{aff82}}
\and W.~J.~Percival\orcid{0000-0002-0644-5727}\inst{\ref{aff83},\ref{aff84},\ref{aff85}}
\and V.~Pettorino\orcid{0000-0002-4203-9320}\inst{\ref{aff81}}
\and S.~Pires\orcid{0000-0002-0249-2104}\inst{\ref{aff86}}
\and G.~Polenta\orcid{0000-0003-4067-9196}\inst{\ref{aff51}}
\and M.~Poncet\inst{\ref{aff87}}
\and L.~A.~Popa\inst{\ref{aff88}}
\and L.~Pozzetti\orcid{0000-0001-7085-0412}\inst{\ref{aff18}}
\and F.~Raison\orcid{0000-0002-7819-6918}\inst{\ref{aff4}}
\and A.~Renzi\orcid{0000-0001-9856-1970}\inst{\ref{aff89},\ref{aff90}}
\and J.~Rhodes\orcid{0000-0002-4485-8549}\inst{\ref{aff22}}
\and G.~Riccio\inst{\ref{aff3}}
\and E.~Romelli\orcid{0000-0003-3069-9222}\inst{\ref{aff30}}
\and M.~Roncarelli\orcid{0000-0001-9587-7822}\inst{\ref{aff18}}
\and E.~Rossetti\orcid{0000-0003-0238-4047}\inst{\ref{aff28}}
\and B.~Rusholme\orcid{0000-0001-7648-4142}\inst{\ref{aff91}}
\and R.~Saglia\orcid{0000-0003-0378-7032}\inst{\ref{aff16},\ref{aff4}}
\and Z.~Sakr\orcid{0000-0002-4823-3757}\inst{\ref{aff92},\ref{aff93},\ref{aff94}}
\and D.~Sapone\orcid{0000-0001-7089-4503}\inst{\ref{aff95}}
\and B.~Sartoris\orcid{0000-0003-1337-5269}\inst{\ref{aff16},\ref{aff30}}
\and M.~Sauvage\orcid{0000-0002-0809-2574}\inst{\ref{aff86}}
\and M.~Schirmer\orcid{0000-0003-2568-9994}\inst{\ref{aff57}}
\and P.~Schneider\orcid{0000-0001-8561-2679}\inst{\ref{aff68}}
\and A.~Secroun\orcid{0000-0003-0505-3710}\inst{\ref{aff59}}
\and G.~Seidel\orcid{0000-0003-2907-353X}\inst{\ref{aff57}}
\and S.~Serrano\orcid{0000-0002-0211-2861}\inst{\ref{aff2},\ref{aff96},\ref{aff1}}
\and E.~Sihvola\orcid{0000-0003-1804-7715}\inst{\ref{aff24}}
\and P.~Simon\inst{\ref{aff68}}
\and C.~Sirignano\orcid{0000-0002-0995-7146}\inst{\ref{aff89},\ref{aff90}}
\and G.~Sirri\orcid{0000-0003-2626-2853}\inst{\ref{aff29}}
\and L.~Stanco\orcid{0000-0002-9706-5104}\inst{\ref{aff90}}
\and J.~Steinwagner\orcid{0000-0001-7443-1047}\inst{\ref{aff4}}
\and P.~Tallada-Cresp\'{i}\orcid{0000-0002-1336-8328}\inst{\ref{aff38},\ref{aff39}}
\and A.~N.~Taylor\inst{\ref{aff36}}
\and I.~Tereno\orcid{0000-0002-4537-6218}\inst{\ref{aff97},\ref{aff98}}
\and N.~Tessore\orcid{0000-0002-9696-7931}\inst{\ref{aff49}}
\and S.~Toft\orcid{0000-0003-3631-7176}\inst{\ref{aff99},\ref{aff100}}
\and R.~Toledo-Moreo\orcid{0000-0002-2997-4859}\inst{\ref{aff101}}
\and F.~Torradeflot\orcid{0000-0003-1160-1517}\inst{\ref{aff39},\ref{aff38}}
\and I.~Tutusaus\orcid{0000-0002-3199-0399}\inst{\ref{aff1},\ref{aff2},\ref{aff93}}
\and L.~Valenziano\orcid{0000-0002-1170-0104}\inst{\ref{aff18},\ref{aff102}}
\and J.~Valiviita\orcid{0000-0001-6225-3693}\inst{\ref{aff61},\ref{aff62}}
\and T.~Vassallo\orcid{0000-0001-6512-6358}\inst{\ref{aff30}}
\and Y.~Wang\orcid{0000-0002-4749-2984}\inst{\ref{aff91}}
\and J.~Weller\orcid{0000-0002-8282-2010}\inst{\ref{aff16},\ref{aff4}}
\and A.~Zacchei\orcid{0000-0003-0396-1192}\inst{\ref{aff30},\ref{aff31}}
\and F.~M.~Zerbi\inst{\ref{aff27}}
\and I.~A.~Zinchenko\orcid{0000-0002-2944-2449}\inst{\ref{aff103}}
\and E.~Zucca\orcid{0000-0002-5845-8132}\inst{\ref{aff18}}
\and J.~Garc\'ia-Bellido\orcid{0000-0002-9370-8360}\inst{\ref{aff104}}
\and M.~Huertas-Company\orcid{0000-0002-1416-8483}\inst{\ref{aff8},\ref{aff105},\ref{aff106}}
\and J.~Macias-Perez\orcid{0000-0002-5385-2763}\inst{\ref{aff107}}
\and J.~Mart\'{i}n-Fleitas\orcid{0000-0002-8594-569X}\inst{\ref{aff108}}
\and V.~Scottez\orcid{0009-0008-3864-940X}\inst{\ref{aff75},\ref{aff109}}}
										   
\institute{Institute of Space Sciences (ICE, CSIC), Campus UAB, Carrer de Can Magrans, s/n, 08193 Barcelona, Spain\label{aff1}
\and
Institut d'Estudis Espacials de Catalunya (IEEC),  Edifici RDIT, Campus UPC, 08860 Castelldefels, Barcelona, Spain\label{aff2}
\and
INAF-Osservatorio Astronomico di Capodimonte, Via Moiariello 16, 80131 Napoli, Italy\label{aff3}
\and
Max Planck Institute for Extraterrestrial Physics, Giessenbachstr. 1, 85748 Garching, Germany\label{aff4}
\and
Dipartimento di Fisica e Astronomia, Universit\`{a} di Firenze, via G. Sansone 1, 50019 Sesto Fiorentino, Firenze, Italy\label{aff5}
\and
University of Trento, Via Sommarive 14, I-38123 Trento, Italy\label{aff6}
\and
INAF-Osservatorio Astrofisico di Arcetri, Largo E. Fermi 5, 50125, Firenze, Italy\label{aff7}
\and
Instituto de Astrof\'{\i}sica de Canarias, E-38205 La Laguna, Tenerife, Spain\label{aff8}
\and
Universidad de La Laguna, Dpto. Astrof\'\i sica, E-38206 La Laguna, Tenerife, Spain\label{aff9}
\and
Institut de Ciencies de l'Espai (IEEC-CSIC), Campus UAB, Carrer de Can Magrans, s/n Cerdanyola del Vall\'es, 08193 Barcelona, Spain\label{aff10}
\and
Department of Mathematics and Physics, Roma Tre University, Via della Vasca Navale 84, 00146 Rome, Italy\label{aff11}
\and
INAF-Osservatorio Astronomico di Roma, Via Frascati 33, 00078 Monteporzio Catone, Italy\label{aff12}
\and
School of Physics, HH Wills Physics Laboratory, University of Bristol, Tyndall Avenue, Bristol, BS8 1TL, UK\label{aff13}
\and
Leiden Observatory, Leiden University, Einsteinweg 55, 2333 CC Leiden, The Netherlands\label{aff14}
\and
INAF-Osservatorio Astronomico di Padova, Via dell'Osservatorio 5, 35122 Padova, Italy\label{aff15}
\and
Universit\"ats-Sternwarte M\"unchen, Fakult\"at f\"ur Physik, Ludwig-Maximilians-Universit\"at M\"unchen, Scheinerstr.~1, 81679 M\"unchen, Germany\label{aff16}
\and
Instituto de F\'isica de Cantabria, Edificio Juan Jord\'a, Avenida de los Castros, 39005 Santander, Spain\label{aff17}
\and
INAF-Osservatorio di Astrofisica e Scienza dello Spazio di Bologna, Via Piero Gobetti 93/3, 40129 Bologna, Italy\label{aff18}
\and
Institute of Theoretical Astrophysics, University of Oslo, P.O. Box 1029 Blindern, 0315 Oslo, Norway\label{aff19}
\and
INAF-Istituto di Astrofisica e Planetologia Spaziali, via del Fosso del Cavaliere, 100, 00100 Roma, Italy\label{aff20}
\and
School of Physics \& Astronomy, University of Southampton, Highfield Campus, Southampton SO17 1BJ, UK\label{aff21}
\and
Jet Propulsion Laboratory, California Institute of Technology, 4800 Oak Grove Drive, Pasadena, CA, 91109, USA\label{aff22}
\and
Department of Physics and Astronomy, University of Utah, 115 South 1400 East, Salt Lake City, UT 84112, USA\label{aff23}
\and
Department of Physics and Helsinki Institute of Physics, Gustaf H\"allstr\"omin katu 2, University of Helsinki, 00014 Helsinki, Finland\label{aff24}
\and
Department of Astronomy, University of Geneva, ch. d'Ecogia 16, 1290 Versoix, Switzerland\label{aff25}
\and
ESAC/ESA, Camino Bajo del Castillo, s/n., Urb. Villafranca del Castillo, 28692 Villanueva de la Ca\~nada, Madrid, Spain\label{aff26}
\and
INAF-Osservatorio Astronomico di Brera, Via Brera 28, 20122 Milano, Italy\label{aff27}
\and
Dipartimento di Fisica e Astronomia, Universit\`a di Bologna, Via Gobetti 93/2, 40129 Bologna, Italy\label{aff28}
\and
INFN-Sezione di Bologna, Viale Berti Pichat 6/2, 40127 Bologna, Italy\label{aff29}
\and
INAF-Osservatorio Astronomico di Trieste, Via G. B. Tiepolo 11, 34143 Trieste, Italy\label{aff30}
\and
IFPU, Institute for Fundamental Physics of the Universe, via Beirut 2, 34151 Trieste, Italy\label{aff31}
\and
Department of Physics "E. Pancini", University Federico II, Via Cinthia 6, 80126, Napoli, Italy\label{aff32}
\and
Dipartimento di Fisica, Universit\`a degli Studi di Torino, Via P. Giuria 1, 10125 Torino, Italy\label{aff33}
\and
INFN-Sezione di Torino, Via P. Giuria 1, 10125 Torino, Italy\label{aff34}
\and
INAF-Osservatorio Astrofisico di Torino, Via Osservatorio 20, 10025 Pino Torinese (TO), Italy\label{aff35}
\and
Institute for Astronomy, University of Edinburgh, Royal Observatory, Blackford Hill, Edinburgh EH9 3HJ, UK\label{aff36}
\and
INAF-IASF Milano, Via Alfonso Corti 12, 20133 Milano, Italy\label{aff37}
\and
Centro de Investigaciones Energ\'eticas, Medioambientales y Tecnol\'ogicas (CIEMAT), Avenida Complutense 40, 28040 Madrid, Spain\label{aff38}
\and
Port d'Informaci\'{o} Cient\'{i}fica, Campus UAB, C. Albareda s/n, 08193 Bellaterra (Barcelona), Spain\label{aff39}
\and
INFN section of Naples, Via Cinthia 6, 80126, Napoli, Italy\label{aff40}
\and
Institute for Astronomy, University of Hawaii, 2680 Woodlawn Drive, Honolulu, HI 96822, USA\label{aff41}
\and
Dipartimento di Fisica e Astronomia "Augusto Righi" - Alma Mater Studiorum Universit\`a di Bologna, Viale Berti Pichat 6/2, 40127 Bologna, Italy\label{aff42}
\and
Jodrell Bank Centre for Astrophysics, Department of Physics and Astronomy, University of Manchester, Oxford Road, Manchester M13 9PL, UK\label{aff43}
\and
European Space Agency/ESRIN, Largo Galileo Galilei 1, 00044 Frascati, Roma, Italy\label{aff44}
\and
Universit\'e Claude Bernard Lyon 1, CNRS/IN2P3, IP2I Lyon, UMR 5822, Villeurbanne, F-69100, France\label{aff45}
\and
Institut de Ci\`{e}ncies del Cosmos (ICCUB), Universitat de Barcelona (IEEC-UB), Mart\'{i} i Franqu\`{e}s 1, 08028 Barcelona, Spain\label{aff46}
\and
Instituci\'o Catalana de Recerca i Estudis Avan\c{c}ats (ICREA), Passeig de Llu\'{\i}s Companys 23, 08010 Barcelona, Spain\label{aff47}
\and
UCB Lyon 1, CNRS/IN2P3, IUF, IP2I Lyon, 4 rue Enrico Fermi, 69622 Villeurbanne, France\label{aff48}
\and
Mullard Space Science Laboratory, University College London, Holmbury St Mary, Dorking, Surrey RH5 6NT, UK\label{aff49}
\and
Universit\'e Paris-Saclay, CNRS, Institut d'astrophysique spatiale, 91405, Orsay, France\label{aff50}
\and
Space Science Data Center, Italian Space Agency, via del Politecnico snc, 00133 Roma, Italy\label{aff51}
\and
University Observatory, LMU Faculty of Physics, Scheinerstr.~1, 81679 Munich, Germany\label{aff52}
\and
Department of Physics, Lancaster University, Lancaster, LA1 4YB, UK\label{aff53}
\and
Felix Hormuth Engineering, Goethestr. 17, 69181 Leimen, Germany\label{aff54}
\and
Technical University of Denmark, Elektrovej 327, 2800 Kgs. Lyngby, Denmark\label{aff55}
\and
Cosmic Dawn Center (DAWN), Denmark\label{aff56}
\and
Max-Planck-Institut f\"ur Astronomie, K\"onigstuhl 17, 69117 Heidelberg, Germany\label{aff57}
\and
NASA Goddard Space Flight Center, Greenbelt, MD 20771, USA\label{aff58}
\and
Aix-Marseille Universit\'e, CNRS/IN2P3, CPPM, Marseille, France\label{aff59}
\and
Universit\'e de Gen\`eve, D\'epartement de Physique Th\'eorique and Centre for Astroparticle Physics, 24 quai Ernest-Ansermet, CH-1211 Gen\`eve 4, Switzerland\label{aff60}
\and
Department of Physics, P.O. Box 64, University of Helsinki, 00014 Helsinki, Finland\label{aff61}
\and
Helsinki Institute of Physics, Gustaf H{\"a}llstr{\"o}min katu 2, University of Helsinki, 00014 Helsinki, Finland\label{aff62}
\and
Laboratoire d'etude de l'Univers et des phenomenes eXtremes, Observatoire de Paris, Universit\'e PSL, Sorbonne Universit\'e, CNRS, 92190 Meudon, France\label{aff63}
\and
SKAO, Jodrell Bank, Lower Withington, Macclesfield SK11 9FT, UK\label{aff64}
\and
Centre de Calcul de l'IN2P3/CNRS, 21 avenue Pierre de Coubertin 69627 Villeurbanne Cedex, France\label{aff65}
\and
Dipartimento di Fisica "Aldo Pontremoli", Universit\`a degli Studi di Milano, Via Celoria 16, 20133 Milano, Italy\label{aff66}
\and
INFN-Sezione di Milano, Via Celoria 16, 20133 Milano, Italy\label{aff67}
\and
Universit\"at Bonn, Argelander-Institut f\"ur Astronomie, Auf dem H\"ugel 71, 53121 Bonn, Germany\label{aff68}
\and
INFN-Sezione di Roma, Piazzale Aldo Moro, 2 - c/o Dipartimento di Fisica, Edificio G. Marconi, 00185 Roma, Italy\label{aff69}
\and
Aix-Marseille Universit\'e, CNRS, CNES, LAM, Marseille, France\label{aff70}
\and
Dipartimento di Fisica e Astronomia "Augusto Righi" - Alma Mater Studiorum Universit\`a di Bologna, via Piero Gobetti 93/2, 40129 Bologna, Italy\label{aff71}
\and
Department of Physics, Institute for Computational Cosmology, Durham University, South Road, Durham, DH1 3LE, UK\label{aff72}
\and
Universit\'e Paris Cit\'e, CNRS, Astroparticule et Cosmologie, 75013 Paris, France\label{aff73}
\and
CNRS-UCB International Research Laboratory, Centre Pierre Bin\'etruy, IRL2007, CPB-IN2P3, Berkeley, USA\label{aff74}
\and
Institut d'Astrophysique de Paris, 98bis Boulevard Arago, 75014, Paris, France\label{aff75}
\and
Institut d'Astrophysique de Paris, UMR 7095, CNRS, and Sorbonne Universit\'e, 98 bis boulevard Arago, 75014 Paris, France\label{aff76}
\and
Institute of Physics, Laboratory of Astrophysics, Ecole Polytechnique F\'ed\'erale de Lausanne (EPFL), Observatoire de Sauverny, 1290 Versoix, Switzerland\label{aff77}
\and
Telespazio UK S.L. for European Space Agency (ESA), Camino bajo del Castillo, s/n, Urbanizacion Villafranca del Castillo, Villanueva de la Ca\~nada, 28692 Madrid, Spain\label{aff78}
\and
Institut de F\'{i}sica d'Altes Energies (IFAE), The Barcelona Institute of Science and Technology, Campus UAB, 08193 Bellaterra (Barcelona), Spain\label{aff79}
\and
School of Mathematics and Physics, University of Surrey, Guildford, Surrey, GU2 7XH, UK\label{aff80}
\and
European Space Agency/ESTEC, Keplerlaan 1, 2201 AZ Noordwijk, The Netherlands\label{aff81}
\and
DARK, Niels Bohr Institute, University of Copenhagen, Jagtvej 155, 2200 Copenhagen, Denmark\label{aff82}
\and
Waterloo Centre for Astrophysics, University of Waterloo, Waterloo, Ontario N2L 3G1, Canada\label{aff83}
\and
Department of Physics and Astronomy, University of Waterloo, Waterloo, Ontario N2L 3G1, Canada\label{aff84}
\and
Perimeter Institute for Theoretical Physics, Waterloo, Ontario N2L 2Y5, Canada\label{aff85}
\and
Universit\'e Paris-Saclay, Universit\'e Paris Cit\'e, CEA, CNRS, AIM, 91191, Gif-sur-Yvette, France\label{aff86}
\and
Centre National d'Etudes Spatiales -- Centre spatial de Toulouse, 18 avenue Edouard Belin, 31401 Toulouse Cedex 9, France\label{aff87}
\and
Institute of Space Science, Str. Atomistilor, nr. 409 M\u{a}gurele, Ilfov, 077125, Romania\label{aff88}
\and
Dipartimento di Fisica e Astronomia "G. Galilei", Universit\`a di Padova, Via Marzolo 8, 35131 Padova, Italy\label{aff89}
\and
INFN-Padova, Via Marzolo 8, 35131 Padova, Italy\label{aff90}
\and
Caltech/IPAC, 1200 E. California Blvd., Pasadena, CA 91125, USA\label{aff91}
\and
Institut f\"ur Theoretische Physik, University of Heidelberg, Philosophenweg 16, 69120 Heidelberg, Germany\label{aff92}
\and
Institut de Recherche en Astrophysique et Plan\'etologie (IRAP), Universit\'e de Toulouse, CNRS, UPS, CNES, 14 Av. Edouard Belin, 31400 Toulouse, France\label{aff93}
\and
Universit\'e St Joseph; Faculty of Sciences, Beirut, Lebanon\label{aff94}
\and
Departamento de F\'isica, FCFM, Universidad de Chile, Blanco Encalada 2008, Santiago, Chile\label{aff95}
\and
Satlantis, University Science Park, Sede Bld 48940, Leioa-Bilbao, Spain\label{aff96}
\and
Departamento de F\'isica, Faculdade de Ci\^encias, Universidade de Lisboa, Edif\'icio C8, Campo Grande, PT1749-016 Lisboa, Portugal\label{aff97}
\and
Instituto de Astrof\'isica e Ci\^encias do Espa\c{c}o, Faculdade de Ci\^encias, Universidade de Lisboa, Tapada da Ajuda, 1349-018 Lisboa, Portugal\label{aff98}
\and
Cosmic Dawn Center (DAWN)\label{aff99}
\and
Niels Bohr Institute, University of Copenhagen, Jagtvej 128, 2200 Copenhagen, Denmark\label{aff100}
\and
Universidad Polit\'ecnica de Cartagena, Departamento de Electr\'onica y Tecnolog\'ia de Computadoras,  Plaza del Hospital 1, 30202 Cartagena, Spain\label{aff101}
\and
INFN-Bologna, Via Irnerio 46, 40126 Bologna, Italy\label{aff102}
\and
Astronomisches Rechen-Institut, Zentrum f\"ur Astronomie der Universit\"at Heidelberg, M\"onchhofstr. 12-14, 69120 Heidelberg, Germany\label{aff103}
\and
Instituto de F\'isica Te\'orica UAM-CSIC, Campus de Cantoblanco, 28049 Madrid, Spain\label{aff104}
\and
Universit\'e PSL, Observatoire de Paris, Sorbonne Universit\'e, CNRS, LERMA, 75014, Paris, France\label{aff105}
\and
Universit\'e Paris-Cit\'e, 5 Rue Thomas Mann, 75013, Paris, France\label{aff106}
\and
Univ. Grenoble Alpes, CNRS, Grenoble INP, LPSC-IN2P3, 53, Avenue des Martyrs, 38000, Grenoble, France\label{aff107}
\and
Aurora Technology for European Space Agency (ESA), Camino bajo del Castillo, s/n, Urbanizacion Villafranca del Castillo, Villanueva de la Ca\~nada, 28692 Madrid, Spain\label{aff108}
\and
ICL, Junia, Universit\'e Catholique de Lille, LITL, 59000 Lille, France\label{aff109}}    
%

%
%
\abstract{

Dual active galactic nuclei (AGNs) are expected in hierarchical galaxy evolution models, in which low-mass galaxies merge to build more massive ones. While observational evidence for dual AGNs is growing in massive galaxies, no clear detection has yet been found in the low-mass regime. 
We used photometry and spectroscopy from the first \Euclid Quick Data Release, combined with a collection of multi-wavelength data from the Dark Energy Spectroscopic Instrument (DESI), the LOw-Frequency ARray (LOFAR) high band antenna, and counterparts in X-ray and mid-infrared catalogues to identify dual AGNs at redshift $z \lesssim 1$. Focusing on low-mass galaxies with stellar masses below 10$^{10}$ M$_{\odot}$, we find nine dual AGN candidates with projected separations ranging from $\sim$20 to 51 kpc. We also find 49 dual AGN candidates in more massive galaxies. We derive a dual AGN fraction of 0.1\% for the low-mass galaxies and estimate that these systems likely trace a population of progenitor black hole pairs that may evolve into bound binaries and eventually coalesce, emitting gravitational waves in the LISA band.
These results constitute the first sample of spectroscopically confirmed dual AGN candidates in low-mass galaxies and have important implications for models in which supermassive black holes grow from lower-mass black holes located in low-mass galaxies, as well as for predictions of gravitational waves from low-mass binary black holes.

}
%
%
%
%
   \titlerunning{First dual AGN in low-mass galaxies}
   \authorrunning{Mezcua et al.}
   
   \maketitle
%
%
%
%
   
\section{\label{sc:Intro}Introduction}

Supermassive black holes (SMBHs) are ubiquitous in massive galaxies (stellar mass $\Mstarsun > 10$), from the local to the high-redshift ($z$) Universe. Yet, how they form and grow is still a matter of debate (see review by \citealt{Alexander2025}). The discovery of SMBHs at $z \sim$ 6--10 (e.g. \citealt{Bogdan2024}; \citealt{Maiolino2024}) suggests they grow from lower-mass seed black holes with 100--10$^6$ M$_\odot$ formed in the early Universe (see reviews by \citealt{Volonteri2010}; \citealt{Mezcua2017}; \citealt{Greene2020}). Galaxy mergers are thought to play a significant role by funnelling gas towards galaxy centres, triggering active galactic nucleus (AGN) activity and thus black hole growth (e.g. \citealt{Hopkins2006}; \citealt{Wild2007}). The AGNs at the centres of merging galaxies also evolve during the merger, progressing through a dual and then a binary (gravitationally bound) phase, until they coalesce and emit gravitational waves. Indeed, hundreds to thousands of dual AGNs have been identified in massive galaxies based on optical, infrared, radio, or X-ray diagnostics (see the compilation by \citealt{Pfeifle2024}) and with separations ranging from less than 1 kpc (e.g. \citealt{Komossa2003}; \citealt{Koss2023}) to more than 50 kpc (\citealt{Liu2011}; \citealt{Koss2012}; \citealt{DeRosa,DeRosa2023}; \citealt{Mannucci+22, Mannucci2023, Ciurlo+23, Scialpi+24}; \citealt{Perna+25}; \citealt{Q1-SP072}; \citealt{Fabricius2026}). 

The presence of AGNs in dwarf or low-mass galaxies ($\Mstarsun \lesssim 10$)\footnote{Both the term `dwarf' versus `low-mass' and the stellar mass threshold adopted to define this regime suffer from a lack of consensus in the literature. We adopt here the same definition ($\Mstarsun \lesssim 10$) as in some recent work on dwarf galaxy mergers and AGNs (e.g. \citealt{Manzano2020}; \citealt{Bichanga2024}; \citealt{Erostegui2025}), but refer to our sources as `low-mass' to be conservative.} is increasingly becoming a commonly observed phenomenon (e.g. \citealt{Reines2013}; \citealt{Mezcua2018,Mezcua2019,Mezcua2023,Mezcua2024}; \citealt{Salehirad2022}; \citealt{Siudek2023}; \citealt{MezcuaDominguez2020,MezcuaDominguez2024}; \citealt{Pucha2025}), and dwarf galaxies are often found to be interacting or merging (e.g. \citealt{Paudel2018}). 
Nevertheless, studies on merger-triggered AGN activity in the dwarf galaxy regime are scant and inconclusive (e.g. \citealt{Kaviraj2019}; \citealt{Bichanga2024}; \citealt{Micic2024}; \citealt{Erostegui2025}), and no spectroscopically confirmed dual AGN has yet been found in low-mass galaxies. 

Local dwarf galaxies are expected to host the relics of those early-Universe seed black holes that did not grow into SMBHs (e.g. \citealt{Mezcua2017}; \citealt{Greene2020}; \citealt{Reines2022}). Therefore, understanding whether galaxy mergers can trigger seed black hole growth in low-mass galaxies, and whether dual AGNs form at all and evolve into binary black holes that coalesce, can have major implications for models of seed black hole formation and evolution (e.g. \citealt{Deason2014}; \citealt{Mezcua2019Nat}). 

Thanks to its superb photometric sensitivity in the optical and near-infrared regimes, the \Euclid mission offers an exciting opportunity for identifying faint merger signatures in the dwarf galaxy regime and better constraining merging galaxy properties (i.e. stellar masses) when performing spectral energy distribution (SED) fitting (\citealt{Q1-SP014}). In addition, \Euclid's near-infrared spectrometer provides the possibility to identify broad-line AGNs whose optical emission lines are obscured by dust (\citealt{Lamperti2017}), making the observatory ideal for investigating dual AGNs in galaxy mergers. 

In this paper we used the first Quick Data Release (Q1) of the \Euclid mission (\citealt{Q1-TP001}) to identify dual AGNs and galaxy mergers across mass scales, with the focus on the low-mass regime. We identify nine dual AGN candidates in low-mass galaxies, the first sample of its kind. The data analysis is reported in \cref{sc: data}, the results and discussion in \cref{sc: results}, and the conclusions in \cref{sec: conclusions}. We assumed a $\Lambda$ Cold Dark Matter cosmology with \mbox{$H_0 = 70$\,km\,s$^{-1}$\,Mpc$^{-1}$}, \mbox{$\Omega_{\rm m} = 0.3$}, and \mbox{$\Omega_{\Lambda} = 0.7$}.

\section{\label{sc: data} Data and analysis }

The \Euclid mission provides high-spatial-resolution (0\arcsecf18; \citealt{EuclidSkyVIS}) optical imaging with the visible instrument (VIS; \IE covers $\lambda = 0.53$--0.92\,\micron) and low-resolution near-infrared spectroscopy and photometry ($R \sim 450$ for objects with a diameter of \ang{;;0.5}) with the Near-infrared Spectrometer and Photometer \citep[NISP; with \YE, \JE, and \HE covering $\lambda = 0.95$--2.02\,\micron; ][]{Schirmer-EP18,EuclidSkyNISP} of millions of galaxies out to $z = 6$ \citep{EuclidSkyOverview}. The Euclid Wide Survey \citep[EWS;][]{Scaramella-EP1} will cover over 14\,500 $\rm deg^2$ to a depth of $\IE=26.2$  (5$\sigma$ point-like source). The Euclid Deep Survey (EDS) will observe three different areas, the so-called Euclid Deep Fields (EDF-North, EDF-N; EDF-South, EDF-S; and EDF-Fornax, EDF-F) covering over 53\,deg$^2$ to a depth of $\IE=28.2$ \citep{EuclidSkyOverview}.

In this study, we used Q1 data \citep{Q1cite,Q1-TP001}, which constitutes a first visit to the EDFs, covering a total area of 63.1\,deg$^2$ at the depth of the EWS \citep{Q1-TP001}. We used the EDF-N, since it is the only field that overlaps with the Dark Energy Spectroscopic Instrument (DESI) Early Data Release (EDR; \citealt{DESI_EDR}), which provides robust spectroscopic redshifts (see \citealt{Q1-SP027} for details; hereafter \citetalias{Q1-SP027}). Starting from the $1.1 \times 10^{7}$ sources in the \Euclid Q1 EDF-N and after applying the quality cuts specified in Sect. 2.1.1 of \citetalias{Q1-SP027}, a total of 24\,922 sources were found to have a DESI spectroscopic counterpart. Crossmatching was done using a conservative \ang{;;0.5} radius, corresponding to approximately six times the DESI fibre positioning accuracy (\citealt{Schlafly2021}). To estimate the contamination rate of the crossmatch, we computed the number density of \Euclid Q1 and DESI EDR in their overlapping area (12.5\,deg$^2$), finding values of 285,111 objects/deg$^2$ and 9,104 objects/deg$^2$, respectively. Using a blunder or contamination radius of \ang{;;0.5}, we then computed the blunder probability as the number of expected \Euclid objects in the blunder area and derived the number of random matches as this blunder probability per the number of DESI objects in the overlapping area. We found 1,905 random matches, which we used to derive the contamination fraction as the ratio between this number and that of actual matches. We obtain a contamination fraction of 7.6\%, which should be taken as an upper limit as it assumes no correlation whatsoever between the \textit{Euclid} and DESI catalogues (see \cref{sc: resultsdualAGN}).

\subsection{\label{sc: pairs} Spectroscopically confirmed galaxy pairs}
From the parent sample of 24\,922 \Euclid Q1 sources with a DESI spectroscopic counterpart, we selected galaxy pairs based on a spectroscopic $z$ separation of $\Delta z < 0.005$, a projected physical separation of $d \lesssim 50$ kpc\footnote{This upper limit is commonly adopted in dual AGN searches, we note that less conservative works consider up to 100 kpc (see review by, e.g. \citealt{Pfeifle2024}).}, and a redshift range of $0.01 < z < 1$ to exclude high-$z$ quasars and ensure a visual identification of pairs or merger signatures (e.g. \citealt{Erostegui2025}). This yields a sample of 619 galaxy pairs in the EDF-N. Since DESI typically undercounts close galaxy pairs due to fibre collisions, we estimate the missing fraction to be 9-14\% (see \cref{app:DESI_fiber_collision} for details).

\subsection{\label{sc: agn} Active galaxies in \Euclid }

To identify AGNs in the sample of 619 galaxy pairs (\cref{sc: pairs}), we used optical and near-infrared spectroscopy, mid-infrared colours, X-ray, and radio diagnostics. For this, we crossmatched the parent sample of 619 galaxy pairs with the recent catalogue of 229\,779 AGN candidates in the \Euclid Q1 fields of \citetalias{Q1-SP027}, who applied mid-infrared colour diagnostics based on Wide-field Infrared Survey Explorer (WISE) AllWISE photometry, emission-line diagnostics based on DESI spectroscopy, and X-ray diagnostics based on an X-ray counterpart in either the 4XMM-DR14 \citep{Webb2020} or the \textit{Chandra} Source Catalog Release 2 Series \citep[][]{Evans_2024}, identified using the Bayesian algorithm \texttt{NWAY} (\citealt{Salvato_2018}; see \citetalias{Q1-SP003} for details)\footnote{The eROSITA first Data Release (\citealt{Merloni2024}) does not cover EDF-N.}.
For the radio diagnostics, we searched for radio counterparts in the Very Large Array Sky Survey (VLASS; \citealt{Gordon2021}) at 3\,GHz and in the LOw-Frequency ARray (LOFAR; \citealt{vanHaarlem2013}) survey of the EDF-N at 144\,MHz using the high band antenna (HBA; \citealt{Bondi2024}). For VLASS we used a commonly adopted search radius of 5\arcsec. For LOFAR we used the \cite{Bisigello25} catalogue, which includes optical-to-near-infrared counterparts for 99.2\% of the LOFAR sources presented in the catalogue by \cite{Bondi2024}.

We then selected AGNs as those sources satisfying at least one of the following nine criteria: 

(i) A DESI spectral type classification \citep[\texttt{SPECTYPE=QSO},][]{DESI_EDR}. 

For sources classified as galaxies \citep[\texttt{SPECTYPE=GALAXY},][]{DESI_EDR}, and using the emission-line fluxes, widths, and equivalent widths provided by \texttt{FastSpecFit} \citep{2023ascl.soft08005M}:

(ii) The detection of broad H\,$\alpha$, H\,$\beta$, \ion{Mg}{ii}, or \ion{C}{IV} emission lines with a full width at half maximum (FWHM) $\geq$ 1200\,\kms.  

(iii) An AGN classification in either the \ion{N}{ii} ([\ion{O}{III}]$\lambda$5007/H\,$\beta$ vs [\ion{N}{II}]$\lambda$6583/H\,$\alpha$), \ion{S}{ii} ([\ion{O}{III}]$\lambda$5007/H\,$\beta$ vs [\ion{S}{II}]$\lambda$6717,6731/H\,$\alpha$), or \ion{O}{i} ([\ion{O}{III}]$\lambda$5007/H\,$\beta$ vs [\ion{O}{I}]$\lambda$6300) emission-line diagnostic (BPT) diagrams (\citealt{Baldwin1981}; \citealt{Kewley2001,Kewley2006}; \citealt{Kauffmann2003}; \citealt{Schawinski2007}; \citealt{Law2021}), with the composite star-forming (SF)/AGN classification also included for the NII BPT diagram.

(iv) A strong or weak AGN classification in the WHAN diagram (\citealt{CidFernandes2010}). 

(v) An AGN classification in either the BLUE BPT (\citealt{Lamareille2010}) or the KEX (\citealt{ZhangHao2018}) diagnostic diagrams, both of which enable AGN identification at $z \gtrsim 0.5$ when the BPT diagrams are not available. For the BLUE BPT diagram, we again considered the composite SF/AGN classification. 

All the emission lines used in the emission-line diagnostics were required to have a signal-to-noise ratio of $\rm S/N \geq 3$.

(vi) The detection of broad near-infrared emission lines (FWHM $\geq$ 1200\,\kms), using spectroscopy from \Euclid's NISP instrument. Spectra were fitted whenever the S/N of the emission lines was sufficient to ensure a meaningful fit and analysis.
We applied a custom interactive fitting procedure designed to simultaneously model each spectrum with a power-law continuum, a set of emission lines, and Fe emission complexes based on Fe templates generated with \texttt{Cloudy} \citep{Ferland1998}.
The fitting code is based on the Levenberg--Marquardt algorithm implemented in the Python package \texttt{Lmfit} \citep{Newville2016}, which allows flexible and robust parameter optimisation (see \cref{app:Euclidemissionlines}).

(vii) An AGN classification based on the \citet{Assef2018} 75\% completeness and 90\% reliability diagnostics, using the \Euclid WISE AllWISE counterparts (see \citetalias{Q1-SP027} for further details).

(viii) An X-ray excess at 2--10 keV of at least 3$\sigma$ above the expected X-ray emission from X-ray binaries (XRBs; e.g. \citealt{Mezcua2016,Mezcua2018}), with the latter computed from the star formation rate and $\Mstar$ derived from SED fitting (see \cref{sc:sed}) following \cite{Lehmer2010}.

(iv) A radio excess of at least 3$\sigma$ above the expected radio emission from stellar processes (e.g. \citealt{Mezcua2019}; \citealt{Reines2020}; see \cref{app:radiocounterparts}).

\subsection{\label{sc:sed} SED fitting}
The stellar masses and star formation rates of the host galaxies were measured via SED fitting. The photometric data were compiled by \citetalias{Q1-SP027} and include \Euclid photometry (\IE, \YE, \JE, and \HE), optical fluxes in the $ugriz$ bands from the UNIONS surveys \citep{Gwyn2025}, Galaxy Evolution Explorer (GALEX) near- and far-ultraviolet fluxes, and mid-infrared fluxes from WISE bands 1--4 or \textit{Spitzer} IRAC channels 1--2 if WISE was not available. The redshifts come from DESI EDR spectroscopy (see \citetalias{Q1-SP027} for more details). To account for AGN emission in the SED fitting, we used the well-established and computationally efficient SED fitting \texttt{Code Investigating GALaxy Emission} (\texttt{CIGALE}; \citealt{Boquien_2019, Yang_2022}), adopting a delayed star formation history with exponential decrease in combination with the physically motivated AGN model \texttt{SKIRTOR} \citep{Stalevski_2016}. 
Further details on the SED fitting method and setup are available in \cite{Q1-SP014}, who developed a method to quantify the reliability of such SED-derived properties as stellar mass by using mock SEDs built from observed quasars and galaxies. In particular, the stellar mass reliability, $R_{M_\star}$, corresponds to the probability of the difference with the true value being smaller than both 0.5\,dex and the measured uncertainties.
\cite{Q1-SP014} also provide a multiplicative factor to obtain larger and more realistic uncertainties to reach at least 0.68 reliability (1$\sigma$). These updated uncertainties are incorporated in \cref{table1}. 
Based on the available photometry, SED fitting can be performed for 24\,790 out of the 24\,922 \Euclid Q1 sources with a DESI spectroscopic counterpart. Applying a stellar mass cut \mbox{$\Mstarsun \leq 10$} to select low-mass galaxies results in 11\,863 low-mass galaxies and 12\,927 massive galaxies. To further ensure the robustness of the stellar masses of the dual AGN host galaxies, we used the $\chi^2$ and the $R_{M_\star}$ parameters derived in \cite{Q1-SP014} and identified as high-confidence fits those sources with reduced $\chi^2 \leq 10$ and $R_{M_\star} > 0.5$. 
This reduced $\chi^2$ threshold has been empirically established to eliminate SED fits with spurious or uncertain photometry (\citealt{Q1-SP014}).
Two examples of the SED fitting performed by \texttt{CIGALE} are shown in Fig.~\ref{app:sed}.

\begin{figure}[ht] 
\centering 
\includegraphics[width=0.49\textwidth]{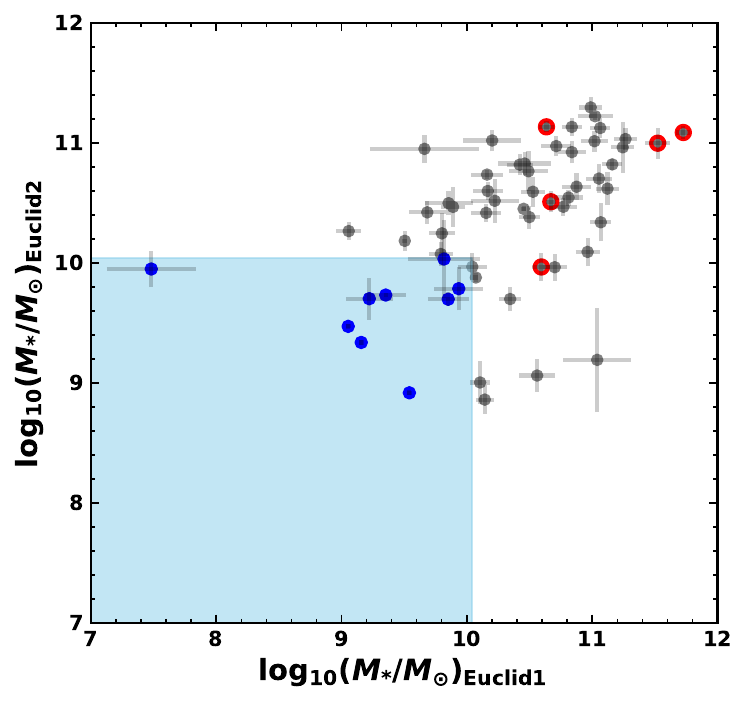} 
\includegraphics[width=0.49\textwidth]{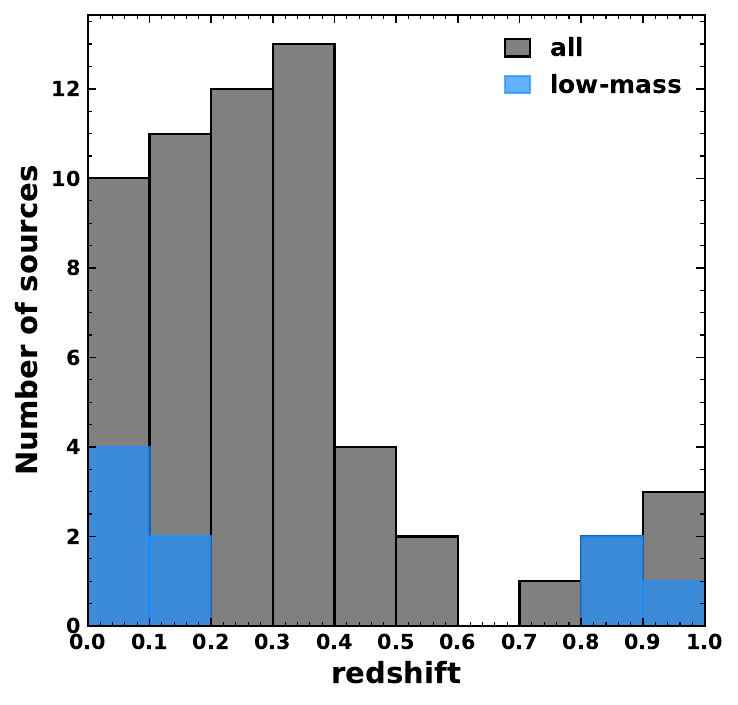} 
\caption{\emph{Top}: Stellar mass for each of the two galaxies (Euclid1 and Euclid2, with indexes 1 and 2 randomly assigned) for the 58 dual AGN candidates. The blue area shows the low-mass regime ($\Mstarsun \leq 10$) used to select dual AGNs in low-mass galaxies (blue dots). All the sources have high-confidence stellar masses based on SED fitting (with $\chi^2 \leq 10$ and $R_{M_\star} > 0.5$; see text) except for the five sources marked in red ($R_{M_\star} < 0.5$).
\emph{Bottom}: Spectroscopic redshift distribution of the 58 dual AGNs (grey bars) and the nine dual AGNs in low-mass galaxies (blue bars).} 
\label{fig:massesdualAGN} 
\end{figure}

\section{\label{sc: results}  Results and discussion}
\subsection{\label{sc: resultsdualAGN} Spectroscopically confirmed dual AGNs}
\begin{figure}[ht] 
\centering 
\includegraphics[width=0.49\textwidth]{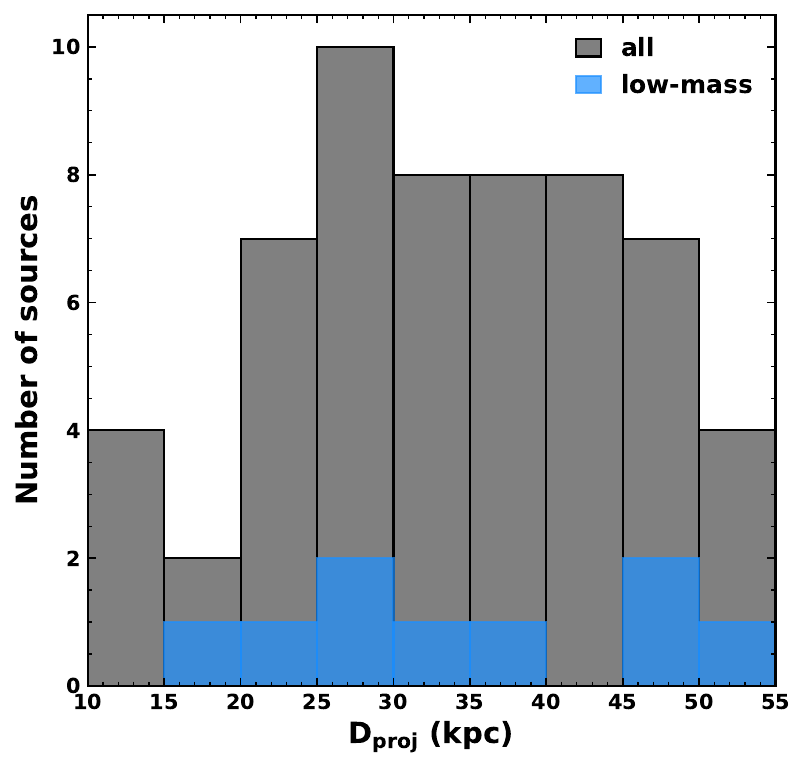} 
\caption{Projected distance distribution of the 58 dual AGN candidates (grey bars) and the nine dual AGNs in low-mass galaxies (blue bars).} 
\label{fig:dproj} 
\end{figure}

In 58 out of the 619 galaxy pairs, each of the two galaxies is found to host an AGN based on at least one of the selection criteria described in \cref{sc: agn}, thus yielding a sample of 58 dual AGN candidates (\cref{fig:massesdualAGN}, top). All the dual AGN systems have SED fits with reduced $\chi^2 \leq 10$ and $R_{M_\star} > 0.5$ except for five systems (marked in red in \cref{fig:massesdualAGN}) with $R_{M_\star} < 0.5$. The spectroscopic redshifts for this sample of 58 dual AGNs range from $z = 0.027$ to 0.950 (Fig.~\ref{fig:massesdualAGN}, bottom), with a gap at $z \sim 0.6$ due to the sources being either DESI Bright Galaxy Survey (whose distribution peaks at $z \sim 0.2$) or Emission Line Galaxy (peak at $z \sim 1.0$) targets (see Fig. 2 in \citealt{DESI_EDR}). 
The projected separation ($D_\mathrm{proj}$) between the two AGNs ranges from $D_\mathrm{proj} =$ 11.2 kpc to 51.8 kpc (\cref{fig:dproj}). In twelve of these systems, the two AGNs are identified by at least two diagnostics and can thus be considered `robust' dual AGNs. The remaining 46 systems, in which one or both AGNs are identified based on only one diagnostic, can be conservatively considered as `candidate' dual AGNs. 

We recomputed the contamination fraction of the \Euclid Q1 and DESI EDR crossmatch adopting a VIS limiting magnitude for \Euclid Q1 of 24.3 mag, which corresponds to the faintest DESI r-band magnitude of the AGN candidates in the dual AGN sample. This yields a \Euclid number density in the overlapping area between \Euclid Q1 and DESI EDR of 70\,542 objects/deg$^2$ and a contamination fraction of 1.9\%, which implies that one out of the 58 dual AGN systems could be a mismatch. Yet, an individual visual inspection of the \Euclid VIS images and their DESI spectroscopic counterparts confirms they are systems within the 0\arcsecf5 radius used in the \Euclid Q1 and DESI EDR crossmatch. 

\subsection{\label{sc: dualAGN} Dual AGNs in low-mass galaxies}

Nine of the 58 dual AGN candidates have \mbox{$\Mstarsun \leq 10$}, yielding a final sample of nine dual AGNs in low-mass galaxies (see \cref{fig:mosaic}). All of the 18 AGN host galaxies have SED fits with reduced $\chi^2 \leq 7$ and $R_{M_\star}>0.5$. Of these, 17 have $R_{M_\star}\geq0.8$, and 14 have reduced $\chi^2 < 3$. Three of the dual AGN systems are located at $z \simeq 0.9$ and are identified as AGNs by the BLUE and KEX diagrams. The remaining six systems are at $z \lesssim 0.1$ and are classified as AGNs based on optical, infrared, or radio (LOFAR) diagnostics (\cref{table1} reports the criteria for each source). Of the nine dual AGNs in low-mass galaxies, one can be considered `robust' because of the two AGNs classified as such by two diagnostics (see \cref{table1}). The remaining systems are considered `candidates', since they are classified as AGNs by only one diagnostic. The latter is particularly relevant for those sources close to the SF/AGN demarcation lines in the BLUE, KEX, or [SII]-BPT diagrams, such as EUCL\,J175306.12$+645855.3$, whose position in the [SII]-BPT is nearly at the intersection between the SF and the AGN region (see \cref{app:desispectroscopy}).

\begin{table*}[ht]
\centering
\caption{Sample of nine dual AGN candidates in low-mass galaxies.}
\label{table1}
\begin{tabular}{llccccc}
\hline
\noalign{\vskip 2pt}
\hline
\noalign{\vskip 2pt}
Euclid1 & Euclid2 & $z_\mathrm{1}$ & $z_\mathrm{2}$ & $D_\mathrm{proj}$ & $\Mstarsun_\mathrm{1}$ & $\Mstarsun_\mathrm{2}$ \\
        &         &         &       &    [kpc]          &         &     \\
\hline
\noalign{\vskip 2pt}
EUCL\,J175226.38$+662659.0$ & EUCL\,J175228.85$+662646.8$ & 0.052 & 0.052 & 19.5 & $9.16 \pm 0.02$ & $9.34 \pm 0.02$ \\
radio                       & radio                       &       &       &      &                 &                  \\       
\hline
\noalign{\vskip 2pt}
EUCL\,J180150.12$+670229.9$ & EUCL\,J180154.58$+670240.4$ & 0.087 & 0.088 & 46.1 & $9.05 \pm 0.02$ & $9.47 \pm 0.02$ \\
NII-BPT, BLUE, radio                       & radio                       &       &       &      &                 &                  \\   
\hline
\noalign{\vskip 2pt}
EUCL\,J181025.14$+664252.8$ & EUCL\,J181026.79$+664235.5$ & 0.088 & 0.088 & 32.7 & $9.54 \pm 0.02$ & $8.92 \pm 0.02$ \\
radio                       & radio                       &       &       &      &                 &                  \\
\hline
\noalign{\vskip 2pt}
EUCL\,J180219.39$+664532.8$ & EUCL\,J180221.60$+664521.3$ & 0.089 & 0.088 & 28.8 & $9.4 \pm 0.2$ & $9.73 \pm 0.02$  \\
WHAN                       & WHAN                       &       &       &      &                 &                  \\     
\hline
\noalign{\vskip 2pt}
EUCL\,J175306.12$+645855.3$ & EUCL\,J175305.12$+645839.0$ & 0.119 & 0.118 & 37.5 & $7.5 \pm 1.3$ & $9.9 \pm 0.2$  \\
SII-BPT                      & WHAN                       &       &       &      &                 &                  \\ 
\hline
\noalign{\vskip 2pt}
EUCL\,J174727.18$+662623.6$ & EUCL\,J174726.73$+662605.0$ & 0.142 & 0.142 & 47.1 & $9.9 \pm 0.2$ & $9.70 \pm 0.06$  \\
NII-BPT, SII-BPT, broad H\,$\alpha$, & radio  &       &       &      &                 &                  \\       
broad Pa\,$\beta$, BLUE, KEX,                       &   &       &       &      &                 &                  \\       
WISE, radio                      &   &       &       &      &                 &                  \\       
\hline
\noalign{\vskip 2pt}
EUCL\,J175759.76$+654524.6$ & EUCL\,J175759.89$+654521.9$ & 0.876 & 0.875 & 21.2 & $9.9 \pm 0.2$ & $9.8 \pm 0.2$ \\
BLUE, KEX                   & BLUE, KEX                        &       &       &      &                 &                  \\   
\hline
\noalign{\vskip 2pt}
EUCL\,J175256.05$+653844.8$ & EUCL\,J175257.08$+653846.1$ & 0.893 & 0.894 & 50.9 & $9.2 \pm 0.3$ & $9.7 \pm 0.2$ \\
KEX                   & BLUE, KEX                        &       &       &      &                 &                  \\  
\hline
\noalign{\vskip 2pt}
EUCL\,J174816.82$+632750.0$ & EUCL\,J174817.34$+632751.2$ & 0.933 & 0.931 & 29.0 & $9.8 \pm 0.5$ & $10.0 \pm 0.5$ \\
KEX                   & BLUE, KEX                        &       &       &      &                 &                  \\       
\hline
\end{tabular}
\tablefoot{
The second row under each \Euclid ID indicates the diagnostic that classifies the source as an AGN. The labels Euclid1 and Euclid2 are randomly assigned.
}
\end{table*}

\subsubsection{Radio counterparts and lack of X-ray detections}
None of the 18 sources in the sample of nine dual AGN candidates in low-mass galaxies have a VLASS counterpart. Eight have a radio counterpart in the LOFAR coverage of the EDF-N with the HBA (\citealt{Bondi2024}), consistent with an AGN origin. 
The LOFAR–\Euclid associations -- while performed using robust maximum likelihood techniques and vetted catalogues (see \cref{app:radiocounterparts}) -- may be particularly challenging in the low-stellar-mass regime.  
In \cref{app:radiocounterparts}, we estimate a LOFAR contamination fraction of 8.3\% based on random match statistics; however, this likely represents a lower limit as the reliability of radio associations degrades at low stellar masses due to weaker radio fluxes and higher source densities (e.g. \citealt{Kondapally2021}; \Euclid Collaboration: Visser et al. in prep). Using a Monte Carlo analysis, \Euclid Collaboration: Visser et al. (in prep) find that the average contamination fraction is $\sim20\%$ in the stellar mass range $9.0 \leq \log_{10}(M_*/M_\odot) \leq 10.0$. 

None of the dual AGNs in low-mass galaxies are found to have X-ray detections in the \textit{Chandra} Source Catalog (they are outside of its footprint) or in 4XMM-DR14. Upper limits (at a 3$\sigma$ significance) on the 0.2-12 keV fluxes can be obtained from XMM Slew and pointed observations\footnote{\url{https://xmmuls.esac.esa.int/upperlimitserver/}} for all the AGNs in the dual AGN low-mass systems. Assuming a power-law photon index $\Gamma =2$, we find upper limits on the 0.2-12 keV X-ray luminosities ranging from 10$^{42}$ erg s$^{-1}$ to 10$^{47}$ erg s$^{-1}$, higher than those typically found in X-ray-emitting AGNs in local low-mass galaxies ($< 10^{42}$ erg s$^{-1}$; e.g. \citealt{Schramm2013}; \citealt{Baldassare2017}; \citealt{Eberhard2025}).

\subsubsection{Black hole masses}
\label{sec: bhmass}
We derived the black hole masses of the dual AGNs in low-mass galaxies using the $M_*$-$M_{\rm BH}$ correlation of \citet{Pucha2025}
for confident broad-line AGN candidates (their Equation 3). We find that the black holes masses are in the range \mbox{$\logten(M_{\rm BH}/M_\odot)$ = 4.0-6.7} with a typical uncertainty of 0.5 dex (see \cref{tableBHmass}).

One of the sources (\object{EUCL\,J174727.18$+$662623.6}) presents broad H\,$\alpha$ emission (see \cref{app:desispectroscopy}), from which \citet{Pucha2025} derive a black hole mass of \mbox{$\logten(M_{\rm BH}/M_\odot) = 7.7 \pm 0.5$} using single-epoch virial techniques. For this source we also fitted the \Euclid NISP spectrum (see \cref{fig:euclidfit}), in particular the Pa\,$\beta$ and HeI emission lines. Both have a FWHM $\geq$ 1200\,\kms, indicative of a broad-line region origin. Using the Pa\,$\beta$ relation (see \cref{app:Euclidemissionlines}) from \cite{LaFranca2015} accounting for the same virial factor ($\epsilon$ = 1) as \citet{Pucha2025}, we obtain a black hole mass of \mbox{$\logten(M_{\rm BH}/M_\odot) = 7.8 \pm 0.5$}, fully consistent with that derived from the broad H\,$\alpha$ emission. For the HeI, we used the relations from \cite{Ricci2017,Ricci2022} and obtain a black hole mass of \mbox{$\logten(M_{\rm BH}/M_\odot) = 7.9 \pm 0.5$} (see eq. \ref{eq3} in \cref{app:Euclidemissionlines}).

We note that for \object{EUCL\,J174727.18$+$662623.6} the black hole mass inferred from the DESI and \Euclid spectra is only marginally consistent with that derived from the $M_*$-$M_{\rm BH}$ calibration of \cite{Pucha2025} within its quoted scatter. This illustrates that individual systems may deviate significantly from simple black hole-galaxy scaling relations and that direct spectroscopic estimates can be important to estimate such masses.

\begin{table*}
\caption{Black hole masses of the dual AGNs in low-mass galaxies derived using the $M_*$-$M_{\rm BH}$ correlation of \citet{Pucha2025}.}
\centering
\label{tableBHmass}
\begin{tabular}{cccc}
\hline \hline
Euclid1 & Euclid2 & $\logten(M_{\rm BH}/M_\odot)_\mathrm{1}$ & $\logten(M_{\rm BH}/M_\odot)_\mathrm{2}$ \\
\hline
EUCL\,J175226.38$+662659.0$ & EUCL\,J175228.85$+662646.8$ & $5.7 \pm 0.4$ & $5.9 \pm 0.4$ \\
EUCL\,J180150.12$+670229.9$ & EUCL\,J180154.58$+670240.4$ & $5.6 \pm 0.4$ & $6.1 \pm 0.4$ \\
EUCL\,J181025.14$+664252.8$ & EUCL\,J181026.79$+664235.5$ & $6.1 \pm 0.4$ & $5.5 \pm 0.4$ \\
EUCL\,J180219.39$+664532.8$ & EUCL\,J180221.60$+664521.3$ & $5.9 \pm 0.6$ & $6.3 \pm 0.5$ \\
EUCL\,J175306.12$+645855.3$ & EUCL\,J175305.12$+645839.0$ & $4.0 \pm 1.7$ & $6.6 \pm 0.6$ \\
EUCL\,J174727.18$+662623.6$ & EUCL\,J174726.73$+662605.0$ & $7.8 \pm 0.5$$^{\dagger}$ & $6.3 \pm 0.5$ \\
EUCL\,J175759.76$+654524.6$ & EUCL\,J175759.89$+654521.9$ & $6.6 \pm 0.6$ & $6.4 \pm 0.6$ \\
EUCL\,J175256.05$+653844.8$ & EUCL\,J175257.08$+653846.1$ & $5.8 \pm 0.7$ & $6.3 \pm 0.6$ \\
EUCL\,J174816.82$+632750.0$ & EUCL\,J174817.34$+632751.2$ & $6.4 \pm 0.9$ & $6.7 \pm 0.9$ \\
\hline
\end{tabular}
\tablefoot{$^{\dagger}$ Black hole mass derived from the broad Pa\,$\beta$ line.}
\end{table*}

\subsubsection{Merger stage}
\begin{figure*}[ht] 
\centering 
\includegraphics[width=0.8\textwidth]{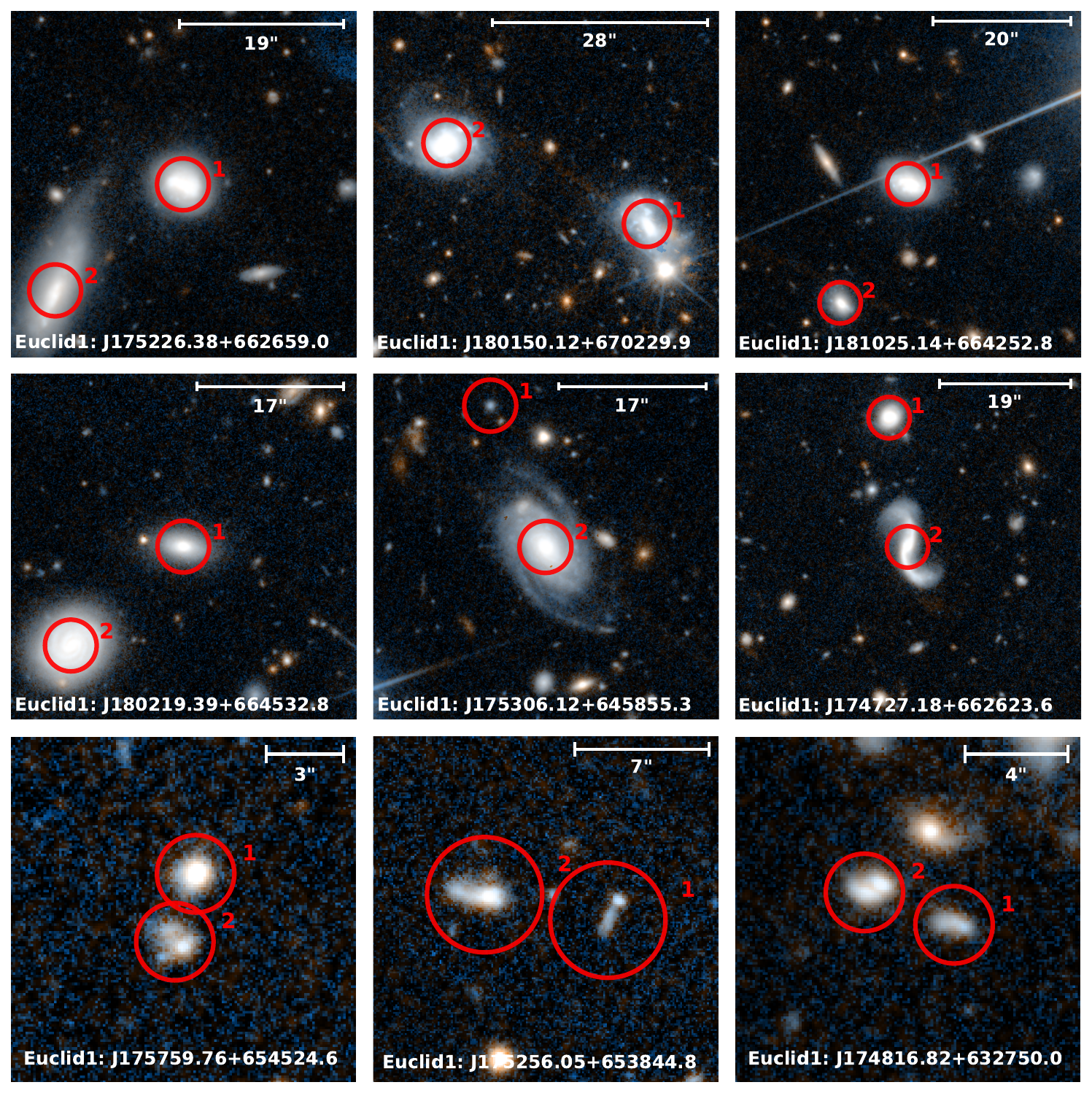} 
\caption{Composite VIS- and NISP-band images of the nine dual AGN candidates in low-mass galaxies sorted in ascending redshift, as in \cref{table1}. The red circles mark the positions of the AGNs. The scale bars show the separation between the two candidate AGNs in arcsec.} 
\label{fig:mosaic} 
\end{figure*}

The $D_\mathrm{proj}$ values between the two AGN candidates in the sample of nine dual AGNs in low-mass galaxies range from 19.5\,kpc to 50.9\,kpc, which suggests they are all either in an early merger stage or have undergone at least a first pass (e.g. \citealt{Erostegui2025}). A visual inspection of the VIS images indeed reveals tidal signatures in some of the sources (e.g. EUCL\,J175226.38$+662659.0$ and EUCL\,J174727.18$+662623.6$ in \cref{fig:mosaic}), while most of the systems seem to be close pairs of galaxies yet to undergo a merger process. This supports previous findings suggesting that AGNs in dwarf galaxies are not necessarily activated by merger events (\citealt{Erostegui2025}).

\subsubsection{Dual AGN fraction}
The sample of nine dual AGNs in low-mass galaxies reported in this paper allows us to derive for the first time a dual AGN fraction in low-mass galaxies. Because of incompleteness effects (the dual AGN candidate sample is based on different surveys and thus different target samples), the fraction should be taken as a lower limit.
We define the dual AGN fraction as the ratio of the dual AGN systems to the total number of low-mass galaxies in the sample (9/11\,863), finding a 0.1\% dual AGN fraction for low-mass galaxies. In the case of massive galaxies, the dual AGN fraction is found to be 0.4\% (49/12\,927). The lower dual AGN fraction for low-mass galaxies is expected given the decline of black hole occupation fraction and of AGN fraction with stellar mass found observationally (e.g. \citealt{Miller2015}; \citealt{Mezcua2018}; \citealt{Birchall2020}; \citealt{Bykov2024}; \citealt{Zou2025}) and in simulations (e.g. \citealt{Bellovary2019}; \citealt{Pacucci2021}).

Although the dual AGN fractions here derived should be taken with caution, they are at least an order of magnitude smaller than those observed in nearby massive systems (e.g. 7.8\% for ultra-hard X-ray-selected AGNs \citealt{Koss2012} and 1.3\% based on optical emission-line diagnostics \citealt{Liu2011}, both at separations $\leq$ 30 kpc), even when considering the total dual AGN fraction of 0.2\%, which accounts for both low-mass and massive galaxies.
Yet, this dual AGN fraction seems consistent with that expected from large-scale hydrodynamical simulations in which the dual AGN fraction (including both low-mass and massive galaxies) is 0.1\% at $z \leq 0.5$ (\citealt{Rosas-Guevara2019}; \citealt{Volonteri2022}; \citealt{Puerto2025}). The dual AGN fraction derived here is also consistent with that from models in which AGN activity is triggered in gas-rich progenitor galaxies when the nuclei of these galaxies are roughly within the half-light radii of their companion galaxies (\citealt{Yu2011}).

\subsubsection{Progenitors of future gravitational wave LISA sources}
To assess the potential relevance of the dual AGNs in low-mass galaxies for future gravitational wave detections as the black holes approach coalescence, we estimated the merger timescales $t_{\rm c}$ from the prescription of \cite{Kitzbichler2008}:

\begin{equation}
	t_{\rm c} = 2.2\,\rm{Gyr} \, \frac{D_{\rm proj}}{50\,kpc} \,\left(\frac{\mathit{M_\star}}{4 \times 10^{10}\, \mathit{h}^{-1}\, \mathit{M_\odot}}\right)^{-0.3} \, \left(1+\frac{\mathit{z}}{8}\right),
\end{equation}
where $M_\star$ is the major stellar mass of the galaxies in the system and $h = 0.7$. This results in $t_{\rm c} = $ 1.9--5.2 Gyr.
The black hole masses derived in \cref{sec: bhmass} imply observed innermost stable circular orbit frequencies $f_\mathrm{ISCO}$ in the interval $\sim$ 0.5--5 mHz, where

\begin{equation}
    f_\mathrm{ISCO} = \frac{c^3}{2 \pi G M_{\rm BH}}\frac{1}{6^{3/2}}  
\end{equation}
and $G$ is the gravitational constant.

Under standard assumptions for the LISA sensitivity curve (LISA Definition Study Report; \citealt{Colpi2024}) and assuming that these pairs evolve into bound black hole binaries and coalesce, their late-inspiral gravitational wave emission could, in principle, be detectable by LISA. We stress, however, that the estimated merger timescales are in some cases longer than the look-back time at the observed redshift, implying that coalescence may occur only at later cosmic times and that AGN activity will not likely remain continuously strong throughout the full merger sequence.

A full reassessment of the space density of such systems and of the corresponding LISA event rate, incorporating these revised black hole masses, merger timescales, and stellar mass functions, is beyond the scope of this work and will be addressed elsewhere. In particular, the relative fraction of low-mass hosts in our sample compared to more massive galaxies should be regarded as a strict lower limit: incompleteness and selection biases, especially against faint and intermediate- to high-redshift systems, almost certainly suppress the observed incidence of such low-mass pairs. 
Consequently, the contribution of the black hole binaries that may eventually descend from these systems to the overall rate of LISA-detectable coalescences will depend sensitively on a careful treatment of incompleteness and selection effects.

\section{Conclusions}\label{sec: conclusions}
In this paper we reported the first sample of nine dual AGNs in low-mass galaxies to date, along with a new sample of 49 dual AGNs in massive galaxies, using \Euclid photometry and spectroscopy, DESI spectroscopy, mid-infrared colours, and LOFAR radio counterparts. The redshifts of the nine dual AGNs in low-mass galaxies range from $\sim$ 0.05 to 0.9, and the AGN-projected separations range from 19.5 to 50.9 kpc. We derive a 0.1\% dual AGN fraction in low-mass galaxies and estimate that these systems likely trace a population of progenitor black-hole pairs that may evolve into bound binaries and eventually coalesce, emitting gravitational waves in the LISA band.

The finding that dual AGNs can form in low-mass galaxies has important implications for seed black hole models, since it provides an avenue for black hole growth if the two black holes coalesce (\citealt{Tamfal2018A}). Thus, these black holes in local dwarf galaxies might not be the unevolved relics of the early-Universe seed black holes (\citealt{Mezcua2019Nat}). The systems identified here can be regarded as possible progenitors of future LISA sources, reinforcing the importance of including dwarf galaxy mergers and dual AGNs in dwarf galaxies in predictions of binary black hole mergers and the gravitational wave background (\citealt{Saeedzadeh2024}). Their contribution to the rate of LISA-detectable black-hole mergers will, however, depend on the time delay between the dual AGN phase and black-hole coalescence, as well as on AGN activity and selection effects.

%
%

\begin{acknowledgements}
\AckEC  
\AckQone
LOFAR data products were provided by the LOFAR Surveys Key Science project (LSKSP; \url{https://lofar-surveys.org/}) and were derived from observations with the International LOFAR Telescope (ILT). LOFAR \citep{vanHaarlem2013} is the Low Frequency Array designed and constructed by ASTRON. It has observing, data processing, and data storage facilities in several countries, which are owned by various parties (each with their own funding sources), and which are collectively operated by the ILT foundation under a joint scientific policy. The efforts of the LSKSP have benefited from funding from the European Research Council, NOVA, NWO, CNRS-INSU, the SURF Co-operative, the UK Science and Technology Funding Council and the Jülich Supercomputing Centre.

M.M. acknowledges support from the Spanish Ministry of Science and Innovation through the project PID2021-124243NBC22. This work was partially supported by the program Unidad de Excelencia Mar\'ia de Maeztu CEX2020-001058-M. 
B.L., J.C., and F.R. acknowledge the support from the INAF Large Grant "AGN \& \Euclid: a close entanglement" Ob. Fu. 01.05.23.01.14.  
A.F. acknowledges the support from project “VLT- MOONS” CRAM 1.05.03.07, INAF Large Grant 2022 “Dual and binary SMBH in the multi-messenger era” Ob. Fu. 1.05.12.01.13, and INAF Mini Grant 2024 “The pc-scale view of HII regions in M33" Ob. Fu. 1.05.24.07.01.
ELSA: Euclid Legacy Science Advanced analysis tools" (Grant Agreement no. 101135203) is funded by the European Union. Views and opinions expressed are however those of the author(s) only and do not necessarily reflect those of the European Union or Innovate UK. Neither the European Union nor the granting authority can be held responsible for them. UK participation is funded through the UK Horizon guarantee scheme under Innovate UK grant 10093177. The authors acknowledge the use of computational resources from the parallel computing cluster of the Open Physics Hub (https://site.unibo.it/openphysicshub/en) at the Physics and Astronomy Department in Bologna.
\end{acknowledgements}

%
%

\bibliography{bibALL} 

%

\begin{appendix}

\section{\label{app:DESI_fiber_collision} DESI fiber collisions}
DESI fibers are limited by the “patrol radius” of the positioners, which results in insufficient available fibers when covering densely populated regions of the sky. Additionally, the physical size of the positioner limits the minimum angular separation between two optical fibers, which means that two (or more) sources with small angular separations cannot be observed simultaneously. Since the combination of these effects is effectively similar to the fiber collisions in SDSS, this issue is also referred to as fiber collisions \citep{Bianchi2025}. A direct consequence of missing closely separated sources is the undercount of physically close pairs, which we estimate here taking an empirical approach. 

First we measure the spectroscopic completeness of DESI EDR in the \Euclid-DESI crossmatch region using two catalogues: the DESI EDR (the observed catalogue) and the Legacy Survey DR9 photometry value-added catalogue of potential targets list (the parent catalogue), which contains all the targets that DESI could have observed \citep{DESI_EDR}. Even after restricting the observed and parent catalogues to the \Euclid-DESI crossmatch region (12.5\,deg$^2$), the large amount of sources in both catalogues (especially the parent) makes it impractical to directly measure the number of pairs as a function of angular separation. Instead, we sample 60 different circular subregions of $\sim$0.061\,deg$^2$, each of which can contain pairs with up to 1000\arcsec\, in angular separation. For each region we derive the number of potential $n_{\rm pot}(\theta_{i})$ and observed $n_{\rm obs}(\theta_{i})$ targets per angular separation bin $\theta_{i}$. We also derive the number $n_{\rm phys}(\theta_{i})$ of actual physical pairs (as defined in \cref{sc: pairs}) per angular bin to calculate the probability that an observed pair is a physical pair as a function of the angular separation as

\begin{equation}
	p_{\rm phys}(\theta_{i}) = \frac{n_{\rm phys}(\theta_{i})}{n_{\rm obs}(\theta_{i})}.
	\label{eq:p_phys}
\end{equation}

Then we use $p_{\rm phys}(\theta_{i})$ to estimate the number of physical missing pairs as

\begin{equation}
	n_{\rm phys\_miss}^{l}(\theta_{i}) = p_{\rm phys}(\theta_{i})\, n_{\rm miss}(\theta_{i}),
	\label{eq:n_phys_low}
\end{equation}
where $n_{\rm miss}(\theta_{i}) = n_{\rm pot}(\theta_{i}) - n_{\rm obs}(\theta_{i})$. We can consider $n_{\rm phys\_miss}^{l}(\theta_{i})$ (and all quantities derived from it) as a lower limit (denoted by the superindex $l$) because $p_{\rm phys}(\theta_{i})$ was obtained assuming that the probabilities of a pair being observed and a pair being physical are independent. However, since physical pairs are more likely to be found in denser regions, they have a lower chance of being observed than random pairs at the same angular separation. This means that $n_{\rm phys\_miss}(\theta_{i})$ might be an underestimate of the actual missing physical pairs. To obtain a more conservative limit, we apply the following correction:
\begin{equation}
	p_{\rm phys}^{\rm corr}(\theta_{i}) = \min\left(1,\frac{p_{\rm phys}(\theta_{i})}{C(\theta_{i})}\right),
	\label{eq:p_phys_corr}
\end{equation}
where $C(\theta_{i}) = n_{\rm obs}(\theta_{i})/n_{\rm pot}(\theta_{i})$ is the pair completeness as a function of angular separation, and we use the $\min()$ function to ensure the probability is never greater than 1. Then
\begin{equation}
	n_{\rm phys\_miss}^{u}(\theta_{i}) = p_{\rm phys}^{\rm corr}(\theta_{i})\, n_{\rm miss}(\theta_{i}).
	\label{eq:n_phys_up}
\end{equation}

\begin{figure}[th] 
\centering 
\includegraphics[width=0.5\textwidth]{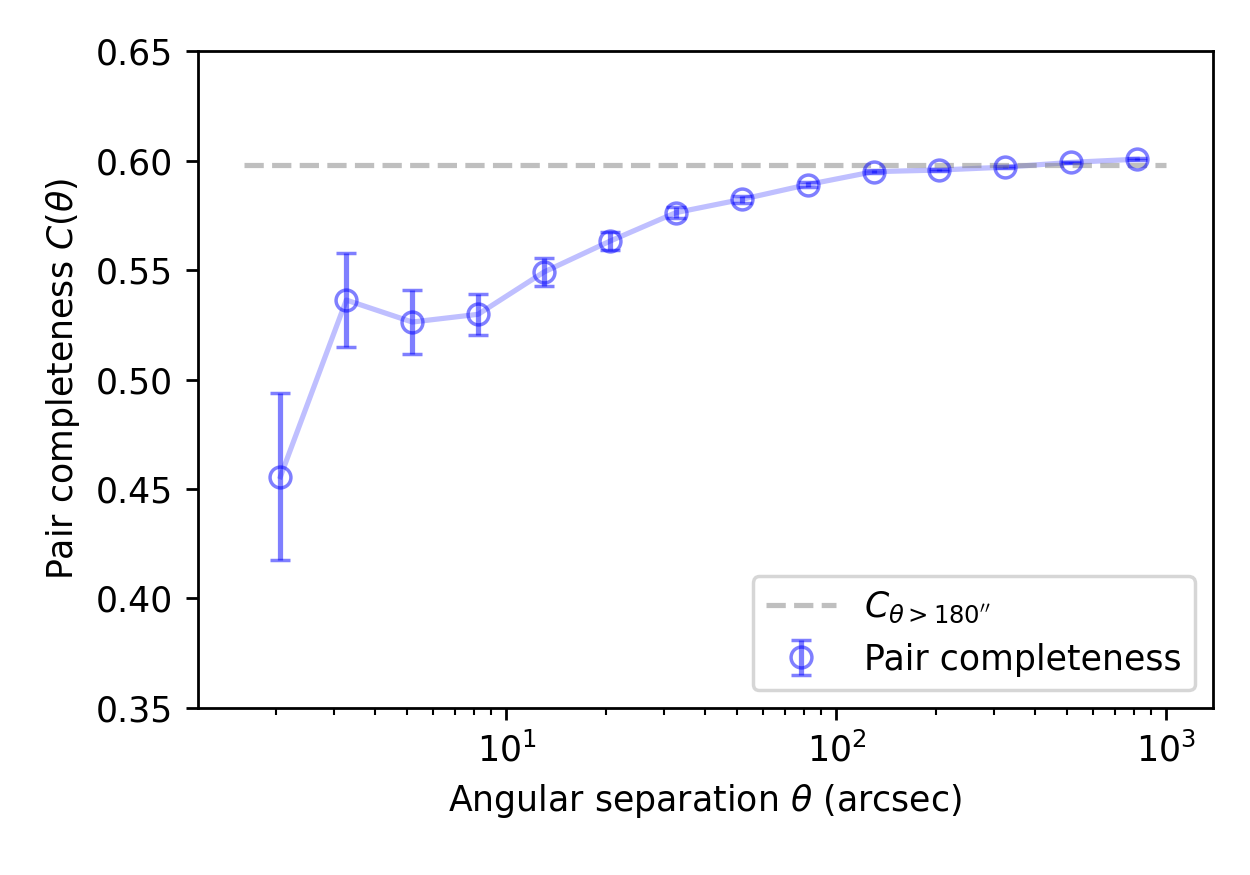} 
\caption{Spectroscopic pair completeness of DESI EDR vs. pair angular separation for the 60 sample regions taken from the DESI-\Euclid crossmatch. The grey dashed line is the average completeness for all pairs with $\theta$>180\arcsec. The vertical bars represent the binomial error of each bin.} 
\label{fig:pair_completeness} 
\end{figure}

After going over all the sample regions, we combine the counts to obtain the spectroscopic pair completeness as a function of angular separation, which is shown in \cref{fig:pair_completeness}. The effect of fiber collision becomes apparent as a decline in completeness from $\sim200\arcsec$ towards smaller separations.

We estimate the relative portion of missing pairs due to fiber collision as

\begin{equation}
	f_{\rm miss\_fc}(\theta_{i}) = C_{\theta>180^{\prime\prime}}-C(\theta_{i}),
	\label{eq:miss_fiber}
\end{equation}
where $C_{\theta>180\arcsec}$ is the average pair completeness for pairs with $\theta$ > 180\arcsec, since at those separations we can expect the fiber collision incompleteness to be negligible (e.g. \citealt{Pinon2025}). We then estimate the number of physical pairs due to fiber collisions as

\begin{equation}
	n_{\rm miss\_fc}^{b}(\theta_{i}) = \frac{f_{\rm miss\_fc}(\theta_{i})}{f_{\rm obs\_miss}(\theta_{i})} \, n_{\rm phys\_miss}^{b}(\theta_{i}),
	\label{eq:n_fiber}
\end{equation}
where $f_{\rm obs\_miss}(\theta_{i}) = 1-C(\theta_{i})$, and the superindex $b$ can be either $u$ or $l$ to get the upper or lower limits, respectively. Finally, we obtain the fraction of missing physical pairs as:

\begin{equation}
	f_{\rm phys\_miss\_fc}^{b} = \frac{1}{C_{\theta>180\arcsec}} \, \frac{\sum_{i=0}^{n_{\rm bin}} n_{\rm miss\_fc}^{b}(\theta_{i})}{\sum_{i=0}^{n_{\rm bin}}[n_{\rm phys\_obs}(\theta_{i}) + n_{\rm phys\_miss}^{b}(\theta_{i})]},
	\label{eq:f_phys_miss}
\end{equation}
where $n_{\rm phys\_obs}(\theta_{i})$ is the number of observed physical pairs per angular bin, $n_{\rm bin}$ is the number of angular separation bins, and we use $C_{\theta>180\arcsec}$ to scale $f_{\rm phys\_miss\_fc}^{b}$ to obtain the missing pair fraction exclusively due to fiber collision.

\section{\label{app:sed} SED fitting}
Two examples of the SED fitting performed by \texttt{CIGALE} are shown in Fig.~\ref{app:sed}.

\begin{figure*}[t!]
\centering
\includegraphics[width=0.67\textwidth]{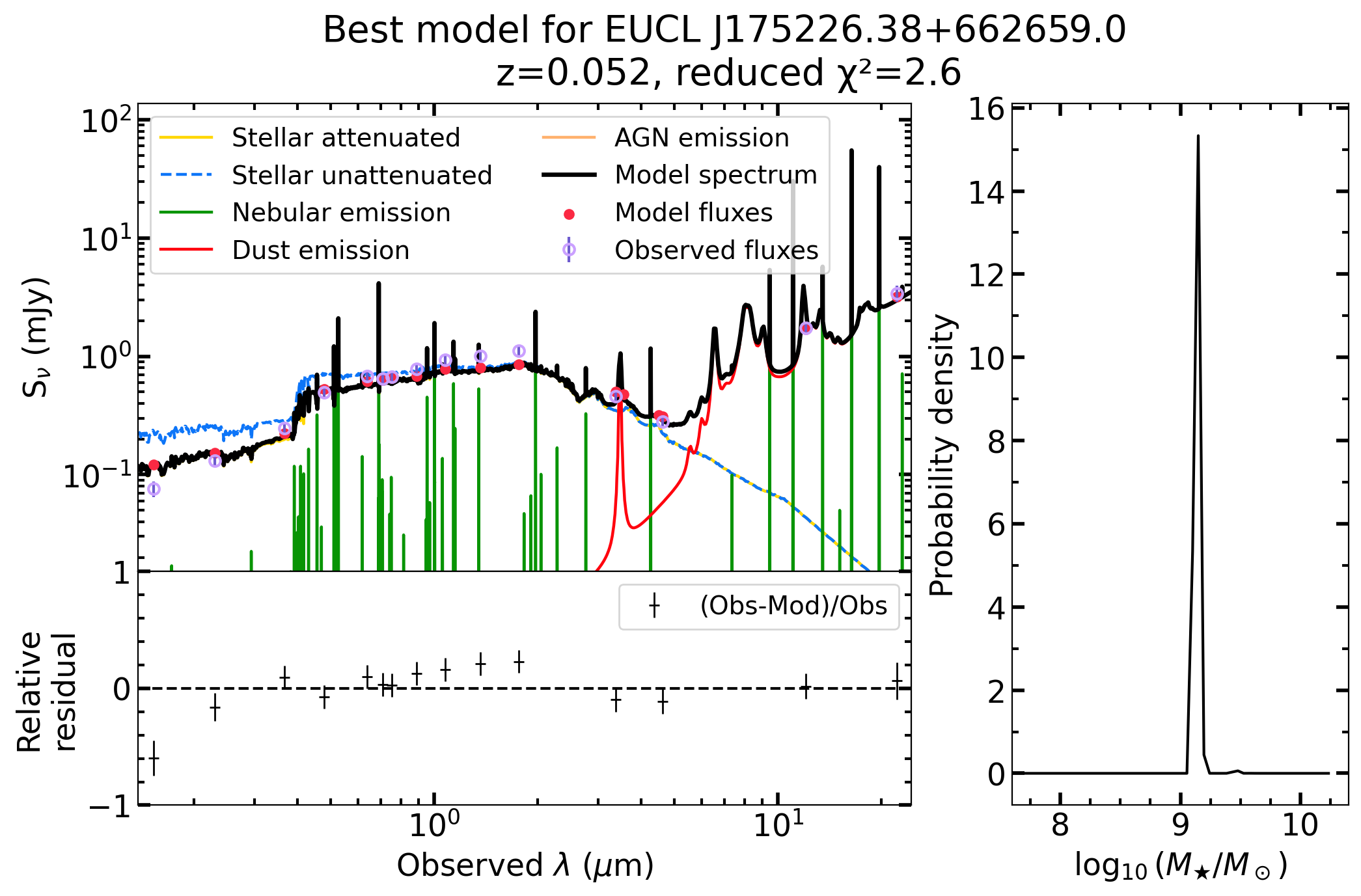}
\includegraphics[width=0.67\textwidth]{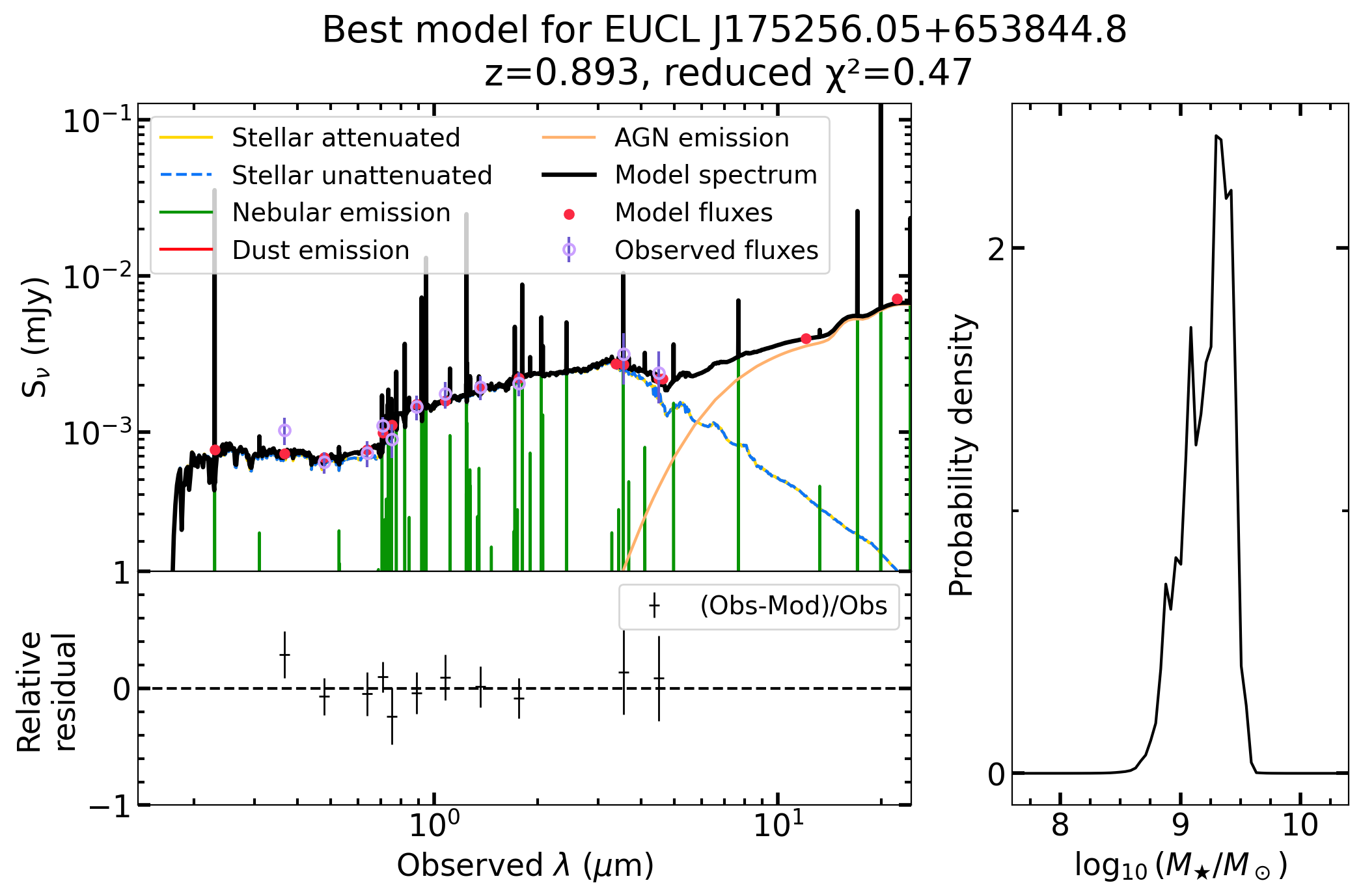}
\caption{Examples of \texttt{CIGALE} best-fit SED models (black line) for one of the galaxies in a low-redshift dual AGN system (\textit{top panel}) and a high-redshift one (\textit{bottom panel}). The observed fluxes in the different bands are shown by purple circles, while the best-fit modelled fluxes in these bands are represented by red dots. The different coloured lines correspond to the various SED components. For each source, a side panel displays the $\Mstar$ probability distribution function returned by \texttt{CIGALE}, showcasing their low-mass nature.}
\label{fig:SEDfit}
\end{figure*}

\begin{table*}[h!]
\centering
\caption{Total and peak radio fluxes, and total radio luminosity at a central frequency of 144\,MHz for those sources in the sample of dual AGN candidates in low-mass galaxies with an AGN radio counterpart. }\label{table2} 
\begin{tabular}{cccc}
\hline
\noalign{\vskip 2pt}
\hline
\noalign{\vskip 2pt}
IAU Name & $S_\mathrm{tot}$ (mJy) & $S_\mathrm{peak}$ (mJy) & $\logten (L_\mathrm{144\,MHz}$/W Hz$^{-1}$) \\
\hline
\noalign{\vskip 2pt}
EUCL\,J174726.73$+662605.0$  &  $0.9 \pm 0.2$ &  $0.33 \pm 0.05$     &      $22.66 \pm 0.08$ \\
EUCL\,J174727.18$+662623.6$  &  $0.8 \pm 0.1$ &  $0.85 \pm 0.04$     &      $22.64 \pm 0.04$ \\
EUCL\,J175226.38$+662659.0$  &  $1.2 \pm 0.1$ &  $0.80 \pm 0.04$     &      $21.90 \pm 0.03$ \\
EUCL\,J175228.85$+662646.8$  &  $0.4 \pm 0.1$ &  $0.22 \pm 0.04$     &      $21.4 \pm 0.1$ \\
EUCL\,J180150.12$+670229.9$  &  $0.7 \pm 0.2$ &  $0.23 \pm 0.04$     &      $22.1 \pm 0.1$ \\
EUCL\,J180154.58$+670240.4$  &  $0.8 \pm 0.1$ &  $0.39 \pm 0.04$     &      $22.16 \pm 0.07$ \\
EUCL\,J181025.14$+664252.8$  &  $1.3 \pm 0.1$ &  $0.92 \pm 0.05$     &      $22.39 \pm 0.04$ \\
EUCL\,J181026.79$+664235.5$  &  $0.6 \pm 0.1$ &  $0.29 \pm 0.05$     &      $22.0 \pm 0.1$ \\
\hline
\end{tabular}
\end{table*}

\begin{figure*}[h!] 
\centering 
\includegraphics[width=0.8\textwidth]{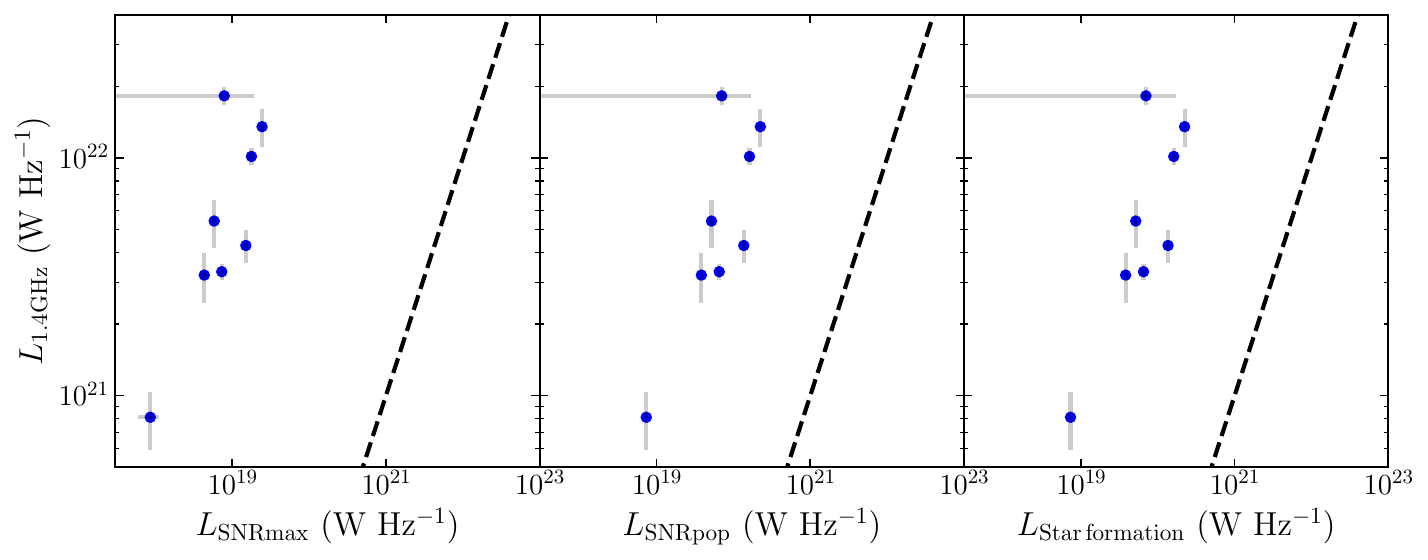} 
\caption{Radio luminosity at 1.4\,GHz of those dual AGN candidates in low-mass galaxies with a LOFAR radio counterpart versus expected radio luminosity from bright SNe and SNRs (\emph{left}), from a population of SNe/SNRs (\emph{middle}), and from star formation (\emph{right}). The dashed lines denote one-to-one correlations. 
} 
\label{fig:radioluminosities} 
\end{figure*}

\section{Radio counterparts}
\label{app:radiocounterparts}
The eight LOFAR sources in the sample of dual AGN candidates in low-mass galaxies are included in the \cite{Bisigello25} catalogue, which includes optical-to-near-infrared counterparts for 99.2\% of the LOFAR sources presented in the catalogue by \cite{Bondi2024}, excluding masked sources. Counterparts are identified using the likelihood ratio method (\citealt{deRuiter1977}; \citealt{Sutherland1992}) based on both the magnitude and colour information, using the optical Subaru Hyper Suprime-Cam (HSC) $i$-band and the 4.5-$\mu$m \textit{Spitzer}/IRAC band. Cross-matches for the most complex sources have been validated with visual inspection by two independent researchers. Given the high success rate of identifications, we directly crossmatch our \Euclid sources with the optical-to-near-infrared positions of the LOFAR counterparts using a matching radius of 1\arcsec. We estimate the contamination fraction of the crossmatch based on the number density of LOFAR in the coverage of the EDF-N (23,309 LOFAR objects over 10\,deg$^2$; \citealt{Bisigello25}) and that of \Euclid EDF-N in that area (285,111 objects/deg$^2$). Using a blunder radius of \ang{;;1.0}, we find 1,611 random matches and derive a contamination fraction of 8.3\%. Yet, all eight \Euclid sources with a LOFAR counterpart are confirmed as the most probable counterpart based on the likelihood ratio analysis presented in \cite{Bisigello25}, with a likelihood ratio at least 100 times larger than the threshold above which a counterpart is considered valid, as derived from the entire LOFAR sample. Using the method described by \cite{Nisbet2018} and \cite{Kondapally2021}, this likelihood ratio corresponds to a probability of a source being a genuine counterpart above 99.4\%.

The LOFAR data were taken at a frequency of 144\,MHz and the image has an angular resolution of 6\arcsec. The root mean square (RMS) of the LOFAR images ranges from 0.04 to 0.06\,mJy\,beam$^{-1}$. The radio source catalogue was extracted by \cite{Bondi2024} using the Python Blob Detection and Source Finder (\texttt{PyBDSF}, \citealt{Mohan2015}). The coordinate-based IAU name and fluxes of the eight radio sources were retrieved from this catalogue and are provided in \cref{table2}. To compute the radio luminosity at 1.4\,GHz and compare it to that expected from stellar processes, we assume a radio spectral index $\alpha$ typical of AGNs, $\alpha = 0.7$ and $S_{\nu} \propto \nu^{-\alpha}$, where $S_{\nu}$ is the radio flux at frequency $\nu$. We use the SED-based star-formation rate of each galaxy to compute the expected contribution at 1.4\,GHz from bright supernovae (SNe) and supernova remnants (SNRs), a population of SNe/SNRs, and from star formation to the radio emission following \cite{Reines2020} and \cite{Flores2025}.
A source is then classified as an AGN when its radio luminosity is at least three times larger than the luminosity predicted from these stellar processes, including their uncertainties (see \cref{fig:radioluminosities}). On the basis of this comparison, we conclude that the radio emission is due to AGN phenomena in all the eight objects.

\begin{figure*}[h] 
\centering 
\includegraphics[width=0.8\textwidth]{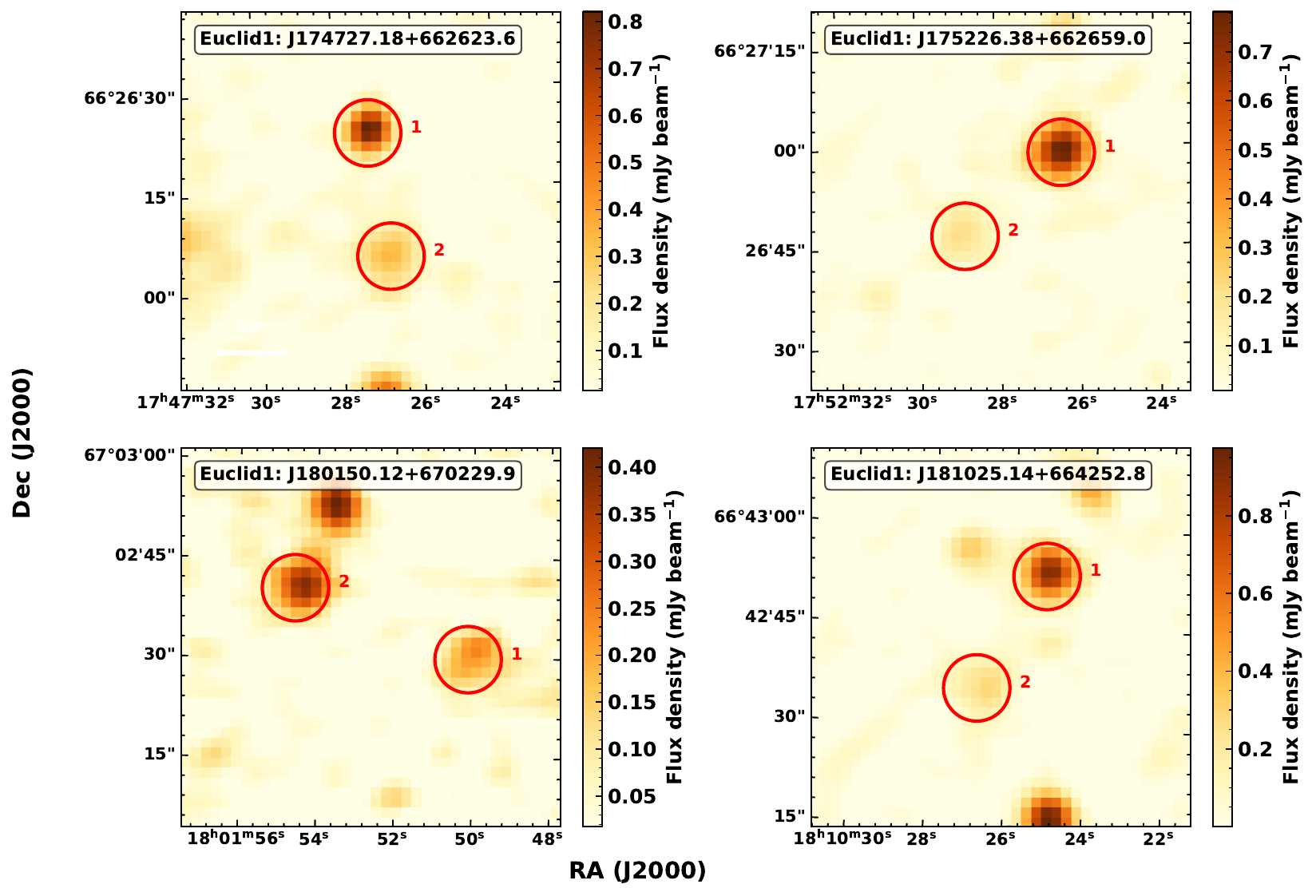} 
\caption{LOFAR EDF-N radio images at 144\,MHz with a resolution of 6\arcsec\, for those dual AGNs in low-mass galaxies with a LOFAR radio counterpart.
The red circles mark the positions of the AGNs, as in \cref{fig:mosaic}.} 
\label{fig:radiomosaic} 
\end{figure*}

\section{\Euclid spectral fitting}
\label{app:Euclidemissionlines}

\begin{figure*}[h]
\centering
\includegraphics[width=0.63\textwidth]{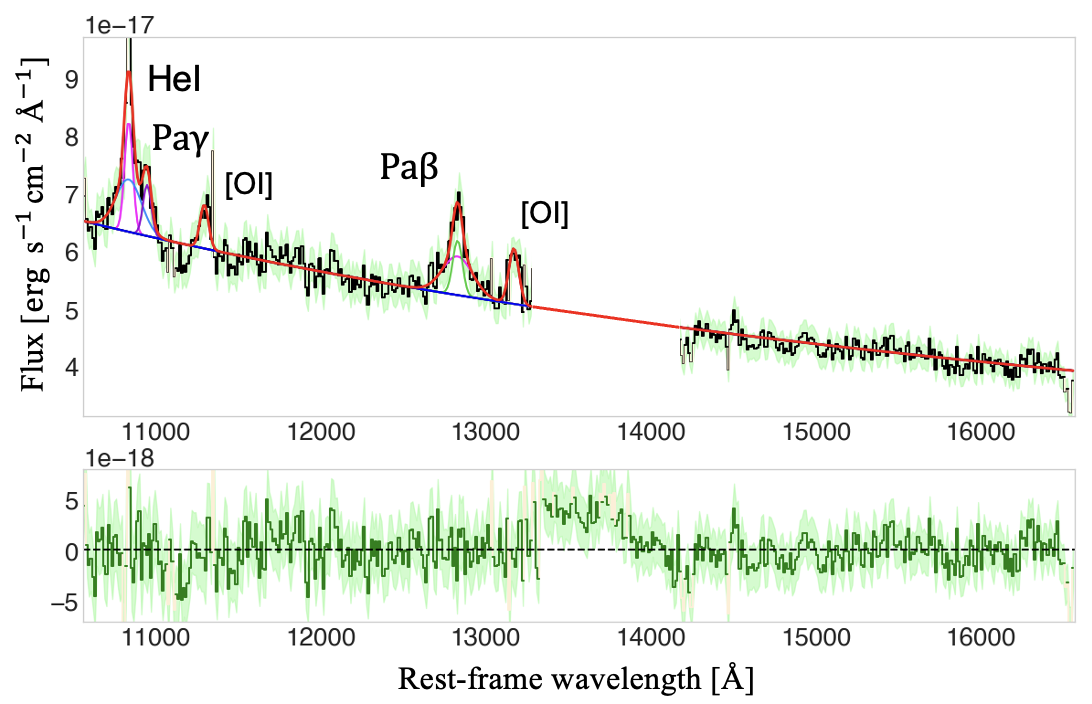}
\caption{
\textit{Euclid} spectrum of \object{EUCL\,J174727.18$+$662623.6}.  
The spectrum is shown in black, with the best-fit model overplotted in red.  Spectral regions affected by instrumental artefacts or elevated noise levels are masked. The fit includes a power-law continuum (in blue) and emission lines, represented in different colours depending on the component. The main lines are labelled in the plot. The bottom panel displays the residuals between the data and the model. The fitting procedure is described in detail in \cref{app:Euclidemissionlines}.
}
\label{fig:euclidfit}
\end{figure*}

In this appendix, we present the results of fitting the \Euclid spectra for the source discussed in \cref{sc: dualAGN}. The spectrum of \object{EUCL\,J174727.18$+$662623.6} is shown in \cref{fig:euclidfit}, with emission lines fitted using the method described in \cref{sc: agn}. We mask regions affected by instrumental features or high noise and we use the DESI spectroscopic redshift to perform the emission line fitting of the \Euclid spectra. 
The observed data are plotted in black, while the best-fit model is overlaid in red. The model comprises a power-law continuum and Gaussian emission lines, including both broad and narrow components as required. Individual fit components are displayed in different colours, with the continuum shown in blue. 
To improve the robustness of the fit, we adopt a physically motivated strategy for the line widths: the narrow lines, which originate from the narrow-line region (NLR), are assumed to have a common velocity dispersion and are therefore fitted with the same fixed width. Similarly, broad lines associated with the broad-line region (BLR) are tied together in width, but allowed to vary independently from the NLR. This approach reduces degeneracies between line fluxes and widths, ensures that the fitted components reflect the expected kinematics of the emitting regions, and prevents unphysical variations in width for lines arising from the same physical region.  
The fluxes and FWHMs of the fitted \Euclid emission lines are listed in \cref{tableEuclidspectroscopy}. We adopt a conservative 20\% uncertainty on the flux measurements of the main emission lines, accounting for both fitting and calibration errors. We fit and report the fluxes and FWHMs only for lines with $\rm S/N \geq 5$.

We compute the black hole mass based on the FWHM and luminosity of the (Pa\,$\beta$)$_\mathrm{b}$ and ({He\,\textsc{i}}~$\lambda\ 1.084$)$_\mathrm{b}$. For Pa\,$\beta$ we use Equation (7) in \cite{LaFranca2015}, adding a $\logten(5.5/4.31)$ term to re-scale the geometrical factor from 4.31 to 5.5 (equivalent to $\epsilon$ = 1):

\begin{align}
\logten\left(\frac{M_{\rm BH}}{M_\odot}\right) &= 7.83 + \logten\left(\frac{5.5}{4.31}\right) \notag \\
&\quad + 0.872\, \left[2 \logten(\mathrm{FWHM}) + 0.5 \logten\left(L_{\mathrm{Pa\,\beta}}\right)\right],
\end{align}
where FWHM is that of (Pa\,$\beta$)$_\mathrm{b}$ in units of 10$^{4}$ km s$^{-1}$ and L$_\mathrm{Pa\,\beta}$ is the (Pa\,$\beta$)$_\mathrm{b}$ line luminosity in units of 10$^{40}$ erg s$^{-1}$.

For HeI we take the relation from \cite{Ricci2017}:

\begin{align}\label{eq1}
\logten\left(\frac{M_{\rm BH}}{M_\odot}\right) &= 7.75 + \logten\left(\frac{5.5}{4.31}\right) \notag \\
&\quad + 2\, \logten(\mathrm{FWHM}) + 0.5 \logten\left(\frac{L_{\mathrm{14\text{-}195\,keV}}}{10^{42}}\right),
\end{align}
where FWHM is that of ({He\,\textsc{i}}~$\lambda\ 1.084$)$_\mathrm{b}$ in units of $10^{4}$ km s$^{-1}$ and $L_\mathrm{14-195\,keV}$ is the 14-195 keV hard-X luminosity in units erg s$^{-1}$. We then re-scale the $L_\mathrm{14-195\,keV}$ to that of HeI using the value reported in \cite{Ricci2022}: 

\begin{equation}\label{eq2}
\logten(L_\mathrm{HeI}) = \logten(L_\mathrm{14-195\,keV})-2.45,
\end{equation}
where $L_\mathrm{HeI}$ is the ({He\,\textsc{i}}~$\lambda\ 1.084$)$_\mathrm{b}$ luminosity in erg s$^{-1}$.

Combining Eqs.(\ref{eq1}) and (\ref{eq2}), we obtain:

\begin{align}\label{eq3}
\logten\left(\frac{M_{\rm BH}}{M_\odot}\right) &= 7.86 + 2\, \logten(\mathrm{FWHM}) \notag \\
&\quad + 0.5\, \left[\logten(L_{\mathrm{HeI}}) - 39.55\right],
\end{align}
where FWHM is that of ({He\,\textsc{i}}~$\lambda\ 1.084$)$_\mathrm{b}$ in units of 10$^{4}$ km s$^{-1}$.

\begin{table}[h]
\centering
\footnotesize
\caption{Fluxes and FWHM values of emission lines detected in EUCL\,J174727.18$+662623.6$.}\label{tableEuclidspectroscopy}
\begin{tabular}{l|cc}
\hline
\noalign{\vskip 2pt}
\hline
\noalign{\vskip 2pt}
Line ID        & Flux & FWHM \\
&    [$10^{-16}$\,erg\,s$^{-1}$\,cm$^{-2}$]   & [km\,s$^{-1}$]   \\      
\hline
\noalign{\vskip 2pt}
({He\,\textsc{i}}~$\lambda\ 1.084$)$_\mathrm{b}$ &  15 $\pm$ 3 & 5083 $\pm$ 1000 \\
({He\,\textsc{i}}~$\lambda\ 1.083$)$_\mathrm{n}$ &  13 $\pm$ 3 & 1469 $\pm$ 290 \\
Pa\,$\gamma$ &  5.0 $\pm$ 1.0 & 1469 $\pm$ 290 \\
{[O\,\textsc{i}]}~$\lambda\,1.129$ &  4.7 $\pm$ 0.9 & 1469 $\pm$ 290 \\
(Pa\,$\beta$)$_\mathrm{b}$ & 16 $\pm$ 3 & 5083 $\pm$ 1000 \\
(Pa\,$\beta$)$_\mathrm{n}$ & 6.5 $\pm$ 1.3 & 1469 $\pm$ 290 \\
\hline
\end{tabular}
\end{table}

\section{\label{app:desispectroscopy} DESI spectroscopy}
\begin{figure*} 
\centering 
\includegraphics[width=0.72\textwidth]{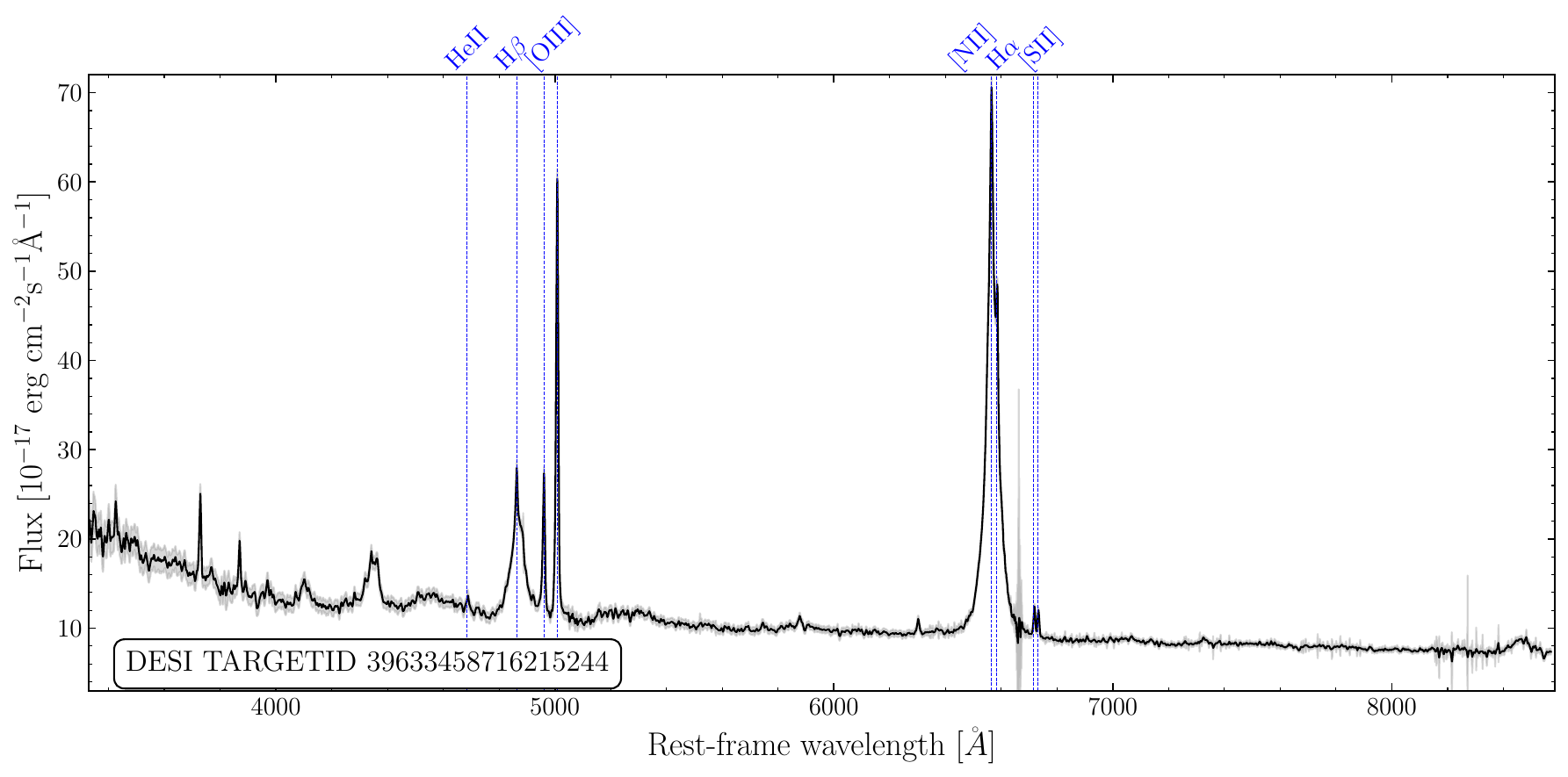} 
\includegraphics[width=0.74\textwidth]{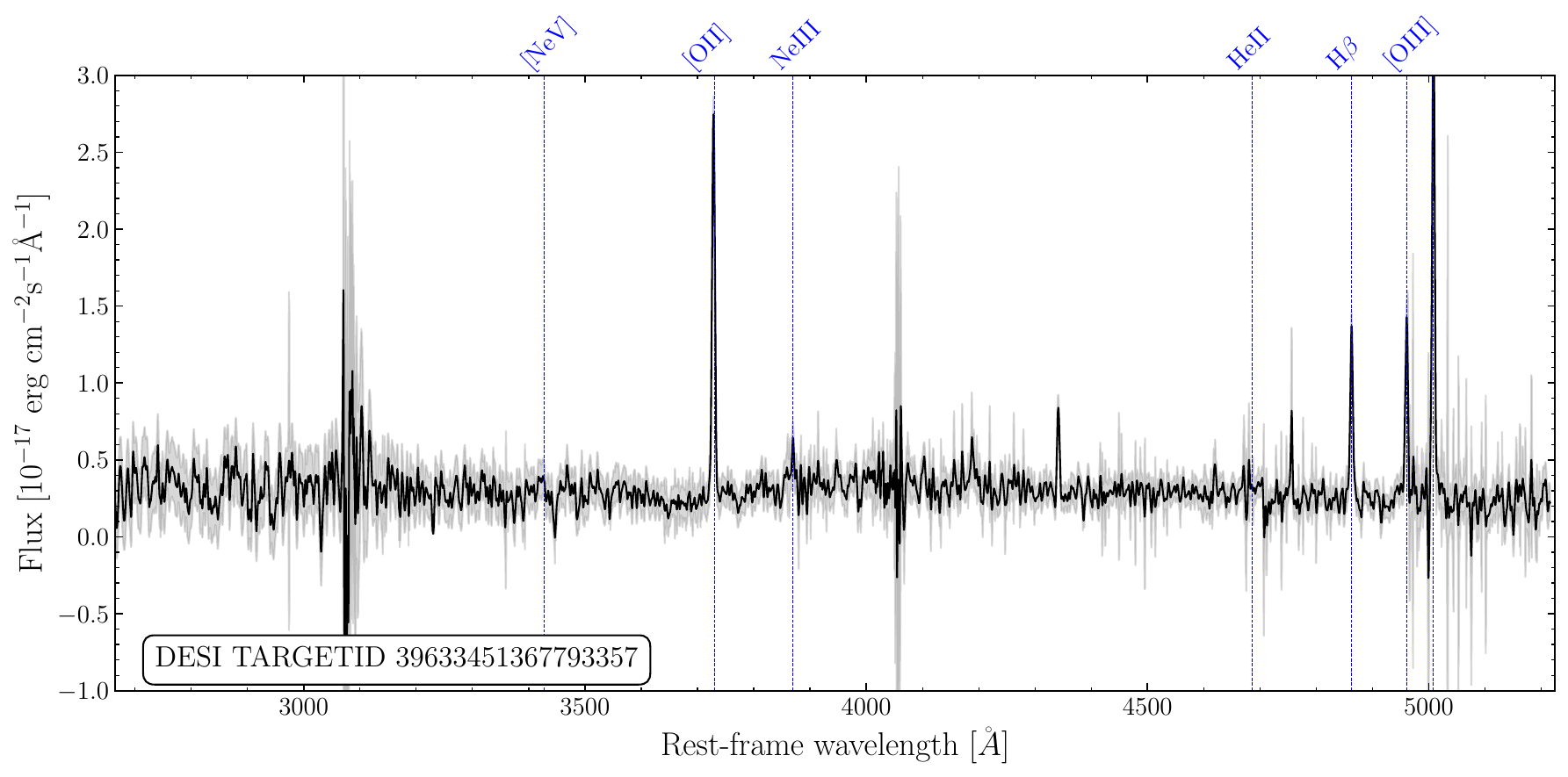} 
\caption{Two examples of the DESI spectra (black line) for dual AGNs in low-mass galaxies: \object{EUCL\,J174727.18$+$662623.6} (\emph{top}), at $z \simeq 0.14$ showing clear broad H\,$\alpha$ emission; and \object{EUCL\,J175759.76$+$654524.6} (\emph{bottom}), at $z \simeq 0.9$ classified as an AGNs based on the BLUE and KEX diagrams. The 1$\sigma$ uncertainty of the flux is shown as a grey line. The blue vertical lines mark the position of those emission lines typically used for AGN classification.} 
\label{fig:DESIexamples} 
\end{figure*}

\begin{figure*} 
\centering 
\includegraphics[width=0.36\textwidth]{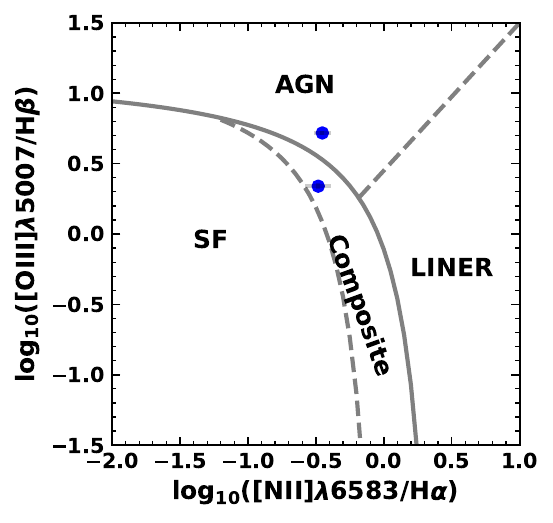} 
\includegraphics[width=0.36\textwidth]{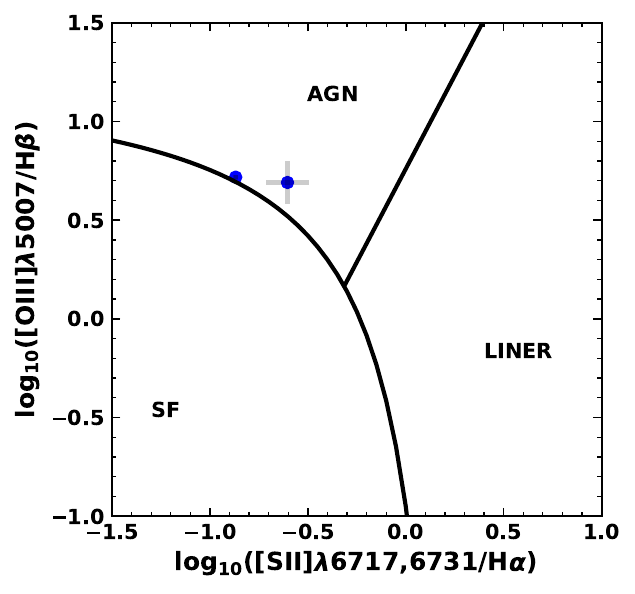} 
\caption{NII-BPT (\emph{left}) and SII-BPT (\emph{right}) diagram  of those sources classified as AGNs in one of these diagnostics based on DESI spectroscopy. In the NII-BPT, the solid line delineates the regions corresponding to AGN and star formation (SF) from \cite{Kewley2001}, while the dashed lines delineate those regions corresponding to SF and composite (combination of AGNs and SF) from \cite{Kauffmann2003}, and those corresponding to AGNs vs low-ionisation nuclear emission-line region (LINER) emission from \cite{Schawinski2007}. In the SII-BPT, the solid lines delineate the regions corresponding to AGNs and SF from \cite{Kewley2001} and those corresponding to AGN vs LINER emission from \cite{Schawinski2007}.}
\label{fig:NIIBPT-SIIBPT} 
\end{figure*}

\begin{figure*} 
\centering 
\includegraphics[width=0.36\textwidth]{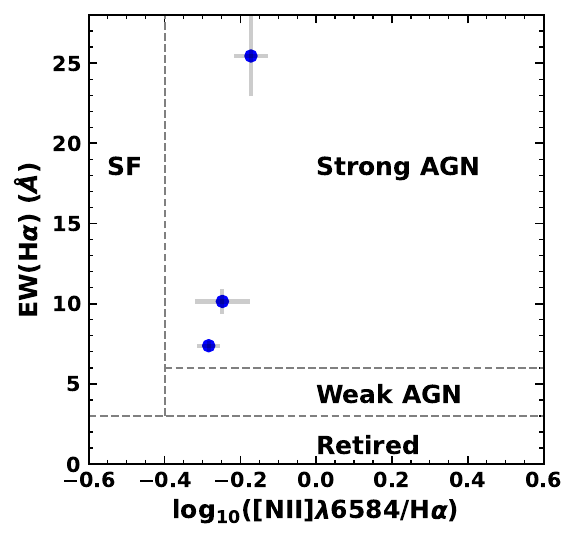} 
\caption{WHAN diagram  of those sources classified as AGNs in this diagnostic based on DESI spectroscopy. The dashed lines separate regions dominated by strong and weak AGN emission, SF emission, and Retired galaxies as defined by \cite{CidFernandes2010}.} 
\label{fig:WHAN} 
\end{figure*}

\begin{figure*}[th] 
\centering 
\includegraphics[width=0.355\textwidth]{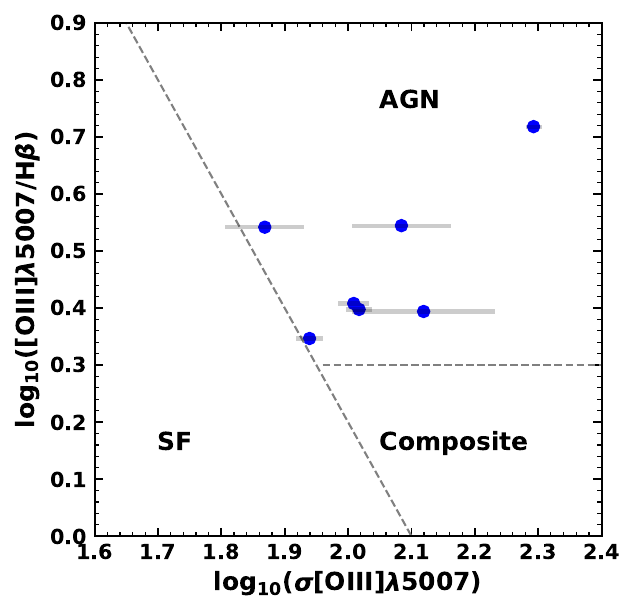} 
\includegraphics[width=0.36\textwidth]{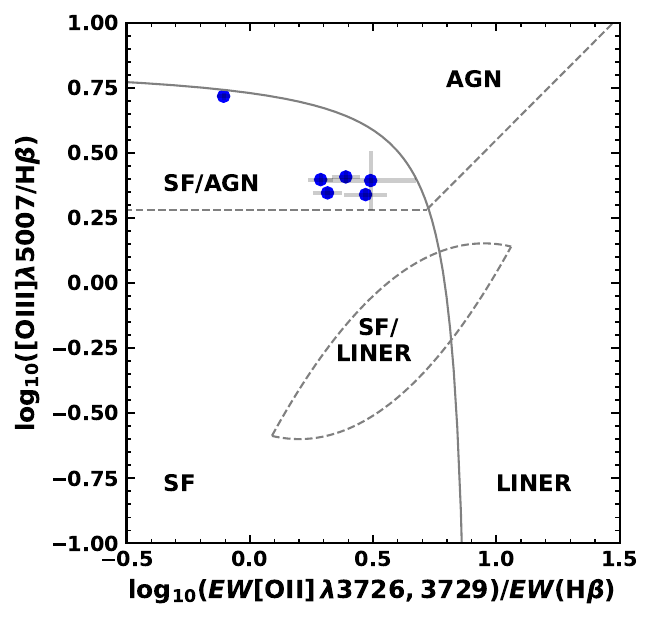} 
\caption{KEX (\emph{left}) and BLUE (\emph{right}) diagrams of those sources classified as AGNs in one of these diagnostics based on DESI spectroscopy. In the KEX diagram, the dashed lines delineate the regions corresponding to AGNs, SF, and composite (combination of AGNs and SF) emission based on \cite{ZhangHao2018}. In the BLUE diagram, the dashed lines indicate regions dominated by AGNs, SF, a combination of SF and AGNs (SF/AGN), LINER, and a combination of SF and LINER (SF/LINER) emission based on \cite{Lamareille2010}.} 
\label{fig:BLUE-KEX} 
\end{figure*}

We visually inspect each individual DESI EDR spectrum of the 18 AGN candidates forming the sample of nine dual AGNs in low-mass galaxies, ensuring that there are no artifacts affecting the emission lines used in the diagnostic diagrams. We show here two examples of the DESI EDR spectra: for the galaxy at $z \simeq 0.14$ with broad H\,$\alpha$ emission (\object{EUCL\,J174727.18$+$662623.6}; \cref{fig:DESIexamples}, top,) and for one of the galaxies at $z \simeq 0.9$ with an AGN classification based on the BLUE and KEX diagrams (\object{EUCL\,J175759.76$+$654524.6}; \cref{fig:DESIexamples}, bottom).
The remaining spectra can be downloaded from the DESI EDR website. 
In \cref{fig:NIIBPT-SIIBPT,fig:WHAN,fig:BLUE-KEX} we also show the NII-BPT, SII-BPT, WHAN, BLUE, and KEX emission-line diagnostics of the sources classified as AGNs based on DESI spectroscopy (see also \cref{table1}).

\end{appendix}

\label{LastPage}
\end{document}